\documentclass[a4paper,11pt]{article}
\pdfoutput=1 

\usepackage{jheppub} 

\usepackage[T1]{fontenc} 

\usepackage[T1]{fontenc} 
\usepackage[utf8]{inputenc} 
\usepackage{microtype} 
\usepackage{lmodern} 
\usepackage{booktabs} 
\usepackage{mdframed} 
\usepackage{graphicx}
\usepackage{scalerel}

\usepackage[backgroundcolor=white]{todonotes}

\newcommand{\be}[1]{ \begin{equation} }
\newcommand{\ee}{\end{equation}}
\newcommand{\bea}[1]{\begin{eqnarray} }
\newcommand{\eea}{\end{eqnarray}}
\newcommand{\bes}{\begin{subequations}}
\newcommand{\ees}{\end{subequations}}

\usepackage{hyperref}
 \usepackage{color}
 \definecolor{dark-red}{rgb}{0.4,0.15,0.15}
 \definecolor{dark-blue}{rgb}{0.15,0.15,0.4}
 \definecolor{medium-blue}{rgb}{0,0,0.5}
 \hypersetup{colorlinks, linkcolor={dark-red}, citecolor={dark-blue}, urlcolor={medium-blue}}

\usepackage{amsmath,amsfonts,amssymb}
\usepackage{eucal} 
\usepackage{mleftright} \mleftright
\usepackage{tensor} 
\usepackage{braket} 
\usepackage{bm} 

\usepackage{mathrsfs} 

\newcommand{\sL}{\mathcal{L}}

\newcommand{\RA}{R^{\scaleto{[-4]}{6pt}}}
\newcommand{\RB}{{R}^{\scaleto{[-2]}{6pt}}}
\newcommand{\RC}{R^{\scaleto{[0]}{6pt}}}
\newcommand{\RD}{R^{\scaleto{[2]}{6pt}}}

\title{\boldmath Galilean fluids from non-relativistic gravity}

\author[1]{Jelle Hartong,}
\author[2]{Aditya Mehra,}
\author[1]{J\o rgen Musaeus}

\affiliation[1]{School of Mathematics and Maxwell Institute for Mathematical Sciences,\\
 University of Edinburgh, Peter Guthrie Tait Road, Edinburgh EH9 3FD, UK}
 \affiliation[2]{School of Basic and Applied Sciences, JSPM University, Gate No. 720, Wagholi, Pune 412207, India}

\emailAdd{j.hartong@ed.ac.uk}
\emailAdd{apm.sbas@jspmuni.ac.in}
\emailAdd{j.s.musaeus@sms.ed.ac.uk}

\abstract{The $1/c$-expansion of general relativity 
appropriately sourced by matter can be used to derive an action principle for Newtonian gravity. The gravitational part of this action is known as non-relativistic gravity (NRG). It is possible to source NRG differently and in such a way that one can construct solutions that are not described by Newtonian gravity (as they do not admit a notion of absolute time). It is possible to include a negative cosmological constant such that NRG admits a non-relativistic AdS solution. This non-relativistic AdS vacuum has Killing vectors that form the Galilean conformal algebra and a boundary that admits a conformal class of Newton-Cartan geometries. This begs the question of whether there exists an analogue of the fluid/gravity correspondence for NRG. In this paper we derive a non-relativistic AdS brane solution of NRG and confirm that it corresponds to the $1/c^2$-expansion of the AdS black brane geometry. We perform a Galilean boost of the non-relativistic AdS brane and derive the associated boundary energy-momentum tensor. We then show that this is the energy-momentum tensor of a massless Galilean fluid and explain how this is linked to the conformal isometries of the boundary. Along the way, we also present several new results for the theory of non-relativistic gravity itself. In particular we present a rewriting that greatly shortens and simplifies the equations of motion of the NRG action.}

\begin{document} 
\maketitle
\flushbottom

\section{Introduction}
The laws of our universe are fundamentally relativistic in nature but in many cases behave approximately non-relativistic. This is true in many areas of physics, not least gravity where the post-Newtonian approximation has been an essential tool in advancing our understanding and predictive ability in general relativity for more than a hundred years. Despite the fact that the non-relativistic regime has been well studied there still appears to be many metaphorical stones left unturned. In fact in recent years, there has been a resurgence of interest in a geometric approach to the non-relativistic regime of general relativity \cite{Tichy:2011te,VandenBleeken:2017rij,Hansen:2019pkl,VandenBleeken:2019gqa,Hansen:2020pqs,Hartong:2023yxo,Hartong:2022lsy,Elbistan:2022plu,Dautcourt:1996pm,Musaeus:2023oyp,Hansen:2020wqw,Hansen:2019vqf,Ergen:2020yop} with the covariant $1/c$-expansion being at the heart of this. This is an off-shell expansion around the limit in which the local lightcone flattens that is described in terms of Newton-Cartan-like geometry. In contrast the post-Newtonian expansion \cite{Blanchet2014,poisson2014gravity} is an explicit weak field and slow-motion expansion around Newtonian gravity, typically based on harmonic gauge. 

One of the key features of the $1/c$-expansion is that it allows for more than Newtonian gravity at leading order and thus opens up to a whole new sector. However, to understand this it is important to have a clear definition of the two sectors first. In the $1/c$-expansion we rewrite the relativistic metric in terms of a timelike vielbein, $T_\mu$, and tensor with $(0,1,\cdots,1)$ signature, $\Pi_{\mu \nu}$, such that
\begin{align}
    g_{\mu \nu} = -c^2T_\mu T_\nu+\Pi_{\mu\nu}\,.
\end{align}
These variables $T_{\mu}$ and $\Pi_{\mu\nu}$ are assumed to be analytic in $1/c$. The key object in terms of defining the different sectors is the leading order term in the expansion\footnote{In this work we do not include odd powers of $1/c$ as these are not needed for our purposes but a more complete treatment should include these as well (see e.g. \cite{Ergen:2020yop,Hartong:2023ckn}).} $T_\mu = \tau_\mu + \mathcal{O}(c^{-2})$. The field $\tau_\mu$ is known as the clock 1-form and tells us about time in our non-relativistic spacetime. The expansion of Einstein's field equations puts a constraint on the clock 1-form 
\begin{align}
    \tau \wedge d \tau=0\,. \label{eq:TTNC}
\end{align}
This means that we can always write $\tau = N dT$ where $N$ and $T$ are scalars. Surprisingly, Einstein's equations still allow for a non-relativistic lapse function that can give rise to gravitational time dilation at leading order. Whether this non-relativistic lapse function is trivial or not is what distinguishes between the two sectors of the expansion. A Newton--Cartan geometry obeying the condition \eqref{eq:TTNC} is known as twistless torsional Newton-Cartan (TTNC) geometry \cite{Christensen:2013lma}.

First, let us consider the so-called weak sector defined by $d \tau=0$. This is the more familiar case where time is absolute. The theory is then described by Newton-Cartan gravity plus corrections thereof. Naturally, this is the sector that contains the post-Newtonian expansion. In fact, the $1/c$-expansion has been used to provide a new framework for the post-Newtonian expansion in previous work \cite{Hartong:2023ckn,companionpaper}. This framework improves upon certain aspects of the otherwise very successful modern approaches, e.g. the Blanchet--Damour approach and the DIRE approach. Most notably, this new framework allows us to perform the post-Newtonian expansion in any gauge with a Newtonian limit\footnote{Not all gauges admit a Newtonian limit since what we usually mean by Newtonian gravity is a gauge fixed version of Newton-Cartan gravity.} and is therefore not limited to the harmonic gauge choice as is the case for the previously mentioned approaches.

The other sector is known as the strong sector and is defined by $\tau\wedge d\tau=0$ with $d \tau \neq 0$. This is a much less understood part of the theory, though it has several interesting features. For example, it has non-relativistic Schwarzschild and Kerr solutions \cite{Ergen:2020yop,Hansen:2020pqs} at leading order which have been shown to pass the three classical tests of GR \cite{Hansen:2020pqs,Hansen:2019vqf}. Another notable feature is the existence of AdS-like solutions which is what led to this project. 
We were motivated by the question of whether or not there exists a non-relativistic analogue of the fluid/gravity correspondence \cite{Bhattacharyya:2007vjd,Baier:2007ix,Rangamani:2009xk,Banerjee:2008th,Haack:2008cp}. In the relativistic case, this is a map between asymptotically AdS black brane solutions (whose parameters such as temperature and velocity are slowly varying functions of the boundary coordinates) of Einstein’s gravity in bulk and fluid dynamics on the boundary. Therefore, in this paper we begin by deriving a non-relativistic AdS brane solution\footnote{We do not refer to this solution as `black' since at this point it is not clear what the causal structure of this solution is.} that corresponds to the $1/c^2$-expansion of the relativistic AdS black brane solution. We perform a Galilean boost of this brane solution in order to introduce a fluid velocity parameter. We then study the conformal boundary of this non-relativistic spacetime and derive an expression for the boundary energy-momentum tensor (EMT) in terms of the bulk parameters. It is shown that this EMT is that of a massless Galilean fluid\footnote{ It should be noted that it is no surprise that we end up with a massless Galilean fluid as it is forced upon us by the conformal symmetry of the boundary.} \cite{Hansen:2020pqs,deBoer:2017ing}. This type of non-relativistic fluid is characterised by a vanishing momentum and therefore differs from the well-known Bargmann (i.e. massive Galilean) fluid which describes many of the fluids we see in everyday life. However, despite the massless Galilean fluids being less well understood it is potentially very interesting in its own right.

The final step in developing this correspondence is to perturb the fluid away from global equilibrium by promoting the parameters of the boosted AdS brane (mass and velocity) to local functions of the boundary coordinates with long wavelength variations. There is a question of regularity in the bulk that needs to be considered which is normally dealt with by going to Eddington–Finkelstein coordinates but due to the absence of a horizon it is not immediately clear what the analogous procedure should be in NRG. We leave such explorations for future work. 

This project focuses on working out the properties of the non-relativistic AdS vacuum (such as its symmetries and definition of the boundary) and how one makes a suitable ansatz for an excited solution such as a non-relativistic AdS brane which is subsequently solved for in the NRG equations of motion. Since NRG is a relatively unexplored theory these first steps already require a fair amount of work. We then furthermore perform the first steps in defining the boundary energy-momentum tensor but leave the programme of holographic renormalisation for NRG for future work. Instead, we rely here on the standard AdS results in Fefferman--Graham gauge and $1/c^2$ expand these.

The work towards establishing a non-relativistic fluid/gravity correspondence serves two main purposes. Firstly, it has relevance to the question of non-relativistic holography. Even though a fluid/gravity correspondence is not enough to claim that there exists a non-relativistic AdS/CFT correspondence, it is a necessary condition for such to be the case. The second and equally important purpose of this paper is to improve our understanding of the strong sector of non-relativistic gravity and expand the non-relativistic gravity (NRG) toolkit. Firstly, it should be noted that the EOM of non-relativistic gravity as derived in \cite{Hansen:2020pqs} (see also \eqref{eq:NRG_EOM1}-\eqref{eq:NRG_EOM4}) have seemed surprisingly long and complicated but in this paper we introduce a rewriting that significantly shrinks and simplifies these EOM as seen in \eqref{eq:diva}-\eqref{eq:NNLO-EOMhh}. This rewriting came about by proving that the process of $1/c^2$ expanding and computing the equations of motion commutes.
The equations were also simplified further by introducing what we call a NLO curvature, a curvature for the NLO fields that transforms nicely under NLO diffeomorphisms. This object also proves useful when deriving the non-relativistic AdS brane solution in section \ref{sec:Excitation&planarSymmetries}. 

NRG consists of two types of fields. The first describe the Newton-Cartan (NC) geometry and the second set form gauge fields on a NC geometry. For the vacuum the latter are zero and for an excited solution they are nonzero. The former set of fields always just describe the non-relativistic AdS spacetime. We show how to make an ansatz for the gauge fields on a non-relativistic AdS spacetime that respects certain symmetries. We do this by defining gauge invariant curvatures and imposing the symmetries first on those. We then use a sequence of gauge fixing operations to solve the conditions on the curvatures in terms of the gauge fields.

As alluded to above, we construct a non-relativistic conformal boundary for spacetimes that are asymptotically locally non-relativistic AdS and show that the corresponding boundary geometry is a type II Newton-Cartan geometry. We do this by suitably modifying the notion of a Penrose boundary to make it applicable to Newton--Cartan geometries.

\subsection{Outline of the paper}
The paper is structured as follows. In section \ref{sec:Non-relativistic gravity} we review the $1/c^2$-expansion of GR to NNLO and introduce new results regarding the theory of non-relativistic gravity. More specifically, in section \ref{sec:2.1} we cover the basics of the $1/c^2$-expansion such as the pre-non-relativistic formulation of GR and non-relativistic gauge transformations. In section \ref{sec:2.2} we introduce the concept of a NLO covariant derivative and the associated curvature that will prove useful later on. Then in section \ref{sec:2.3} we review the derivation of the non-relativistic gravity action and present a new and simpler form of its EOM. In section \ref{sec:VacSym&Bdy} we study the vacuum non-relativistic AdS solution, its symmetries and the construction of a conformal boundary in NRG. Specifically, in section \ref{sec:3.1} we give the non-relativistic AdS solution along with expressions for the LO connection and curvature which will be useful later on. In section \ref{sec:3.2} we compute the Killing symmetries of the non-relativistic AdS vacuum and find that we get a infinite lift of the finite-dimensional Galilean conformal algebra (GCA) \cite{Bagchi:2009my, Bagchi:2009ca, Bagchi:2014ysa, Bagchi:2015qcw, Bagchi:2017yvj, Mehra:2021sfx, Festuccia:2016caf}. Next, in section \ref{sec:3.3} we construct a notion of conformal boundary for spacetimes that are asymptotically locally non-relativistic AdS. We then compute gauge transformation of the boundary fields and show that we get a conformal class of type II Newton-Cartan geometries on the boundary. In section \ref{sec:Excitation&planarSymmetries} we construct the non-relativistic AdS brane solution, perform a Galilean boost in order to introduce velocity and show that the boundary EMT corresponds to that of a massless Galilean fluid. More specifically, in section \ref{sec:4.1} we start with an ansatz that is asymptotically AdS and has planar symmetries. We then solve the NRG EOM for this ansatz and show that we end up with a solution that corresponds to the $1/c^2$-expansion of the AdS black brane. In section \ref{sec:GalBoost} we perform a Galilean boost of the non-relativistic AdS brane solution. Assuming that the $1/c^2$-expansion and holographic renormalisation commute we derive the boundary energy-momentum tensor for the non-relativistic AdS brane. Finally, in section \ref{sec:DualFluid} we show that this EMT corresponds to the EMT of a massless Galilean fluid and explain why this was inevitable due to the conformal symmetry of the boundary. 

We also have several appendices. In appendix \ref{app:curvature} we collect useful formulas and cover basic properties of connections and associated curvatures in the case of non-zero torsion. In appendix \ref{app:EOM} we derive ward identities for the NRG action, show that the NRG EOM are equivalent to the $1/c^2$-expansion of Einstein's field equations and present the details of how to simplify the NRG EOM. In appendix \ref{app:ansatz} we derive, up to a gauge transformation, the most general (spatial components of) $\Phi_{\mu \nu}$ and $m_\mu$ that respect time translation, boundary spatial translation and boundary rotations under the assumption that the leading order geometry is non-relativistic AdS. This is then used as the ansatz in section \ref{sec:Non-relativistic gravity}. Finally, in appendix \ref{App:NRExpansionOfAsympAdS} we $1/c^2$-expand the relativistic AdS black brane metric including its boundary EMT.

\section{Non-relativistic gravity} \label{sec:Non-relativistic gravity}
In this section we review the covariant $1/c^2$-expansion of general relativity much of which is also covered with greater detail in \cite{Hansen:2020pqs,Hartong:2022lsy} . This includes the rewriting of general relativity in terms of pre-non-relativistic (PNR) variables, expansion of the PNR variables and their gauge transformations as well as expanding the Einstein-Hilbert action to next-to-next-to-leading (NNLO) order. This order is generally referred to as non-relativistic gravity and is the order at which one can get Newtonian gravity as a solution. In this section we also cover the definition of a covariant derivative and associated connection that is compatible with the Newton-Cartan metric variables. From this connection one can then define an associated curvature. These are key objects in the theory of non-relativistic gravity. This brings us to the new results of this section, which includes extending the notion of covariant derivatives to objects that transform under NLO diffeomorphisms and from that the notion of a NLO curvature. As we will see later on, this will prove a useful object when constructing solutions to the EOM starting from the desired symmetries of the spacetime. It has also been useful in simplifying the EOM of the NRG action that has previously been untenably long and complicated. Specifically, what before was almost a  page long (see \eqref{eq:NRG_EOM4}) can now be written on two lines (see \eqref{eq:NNLO-EOMhh}). However, the main reason for this simplification did not come from the introduction of the subleading curvature but instead from realising that the EOM of the other fields can be used to greatly simplify \eqref{eq:NRG_EOM4}, as shown by equation \eqref{eq:EOM-Simplification}.

\subsection{Fields and gauge transformations} \label{sec:2.1}

The first step in the covariant $1/c^2$-expansion is to rewrite general relativity in terms of pre-non-relativistic variables with the purpose of preparing the theory for an expansion around a limit in which we have local Galilei rather than Lorentzian symmetry. We start by decomposing the relativistic vielbein $E^A_\mu = (c T_\mu, \mathcal{E}^a_\mu)$ and its inverse $E_A^\mu = (\frac{1}{c} T^\mu, \mathcal{E}_a^\mu)$ with spacetime index $\mu =0,1,\cdots, D-1$, tangent space index $A=0,1\cdots,D-1$ and spatial tangent space index $a=1,\cdots,D-1$. $D$ is the total number of spacetime dimensions. We are then always allowed to make the following decomposition of the metric and its inverse 
\begin{eqnarray}
    g_{\mu\nu} & = & -c^2T_\mu T_\nu+\Pi_{\mu\nu}\,,\\
    g^{\mu\nu} & = & -c^{-2}T^\mu T^\nu+\Pi^{\mu\nu}\,,
\end{eqnarray}
where $T_\mu$ is the timelike vielbein ($T^\mu T_\mu =-1$) while $\Pi_{\mu\nu}=\delta_{ab}\mathcal{E}^a_\mu\mathcal{E}^b_\nu$ and $\Pi^{\mu\nu}=\delta^{ab}\mathcal{E}_a^\mu\mathcal{E}_b^\nu$ are contractions of the spacelike vielbein and their inverse, such that 
\begin{equation}
    \Pi_{\nu\rho}\Pi^{\rho\mu}-T^\mu T_\nu=\delta^\mu_\nu\,. \label{eq:Completeness}
\end{equation}
Using this and the fact that $\Pi_{\mu\nu}$ and $\Pi^{\mu\nu}$ have signature $(0,1,\cdots,1)$ we must also have
\begin{equation}
    T^\mu\Pi_{\mu\nu}=0\,,\qquad T_\mu\Pi^{\mu\nu}=0\,. \label{eq:Orthogonality}
\end{equation}
The basic assumption of the $1/c^2$-expansion is that $T_\mu$ and $\Pi_{\mu \nu}$ are analytic in $c^{-2}$. 

We can then expand the PNR variables in $1/c^2$
\begin{eqnarray}
    T_\mu & = & \tau_\mu+c^{-2}m_\mu+c^{-4}B_\mu+\mathcal{O}(c^{-6})\,,\\
    \Pi_{\mu\nu} & = & h_{\mu\nu}+c^{-2}\Phi_{\mu\nu}+c^{-4}\psi_{\mu\nu}+\mathcal{O}(c^{-6})\,,
\end{eqnarray}
with the expansion of the inverse given by
\begin{eqnarray}
    T^\mu & = & v^\mu+c^{-2}\left(v^\mu v^\rho m_\rho-h^{\mu\rho}v^{\sigma}\Phi_{\rho\sigma}\right)+\mathcal{O}(c^{-4})\,,\\
    \Pi^{\mu\nu} & = & h^{\mu\nu}+c^{-2}\left(-h^{\mu\rho}h^{\nu\sigma}\Phi_{\rho\sigma}+v^\mu h^{\nu\rho}m_\rho+v^\nu h^{\mu\rho}m_\rho\right)+\mathcal{O}(c^{-4})\,.
\end{eqnarray}
Here the expansion of the relations in \eqref{eq:Completeness} and \eqref{eq:Orthogonality} were used to express the subleading corrections to $T^\mu$ and $\Pi^{\mu \nu}$ in terms of the other fields. From \eqref{eq:Orthogonality} it also follows that $T^\mu T^\nu\Pi_{\mu\nu}=0$, which at order $c^{-2}$ implies that 
\begin{equation}
    v^\mu v^\nu\Phi_{\mu\nu}=0\,.
\end{equation}
We will use this in later sections.

It is the leading order fields $(\tau_\mu, h^{\mu \nu})$ as well as their projective inverses $(v^\mu, h_{\mu \nu})$ that describe the geometry of the non-relativistic spacetime while the subleading corrections are interpreted as gauge fields. The fields $(v^\mu, h_{\mu \nu})$ are called projective inverses because as we will see below they are determined up to a local Galilean boost. Moreover, proper time along a curve $\gamma$ is given by the line integral over $\gamma$ of the clock 1-form, $\tau_\mu dx^\mu$.

 In section \ref{sec:GalBoost} and appendix \ref{App:NRExpansionOfAsympAdS} we will also make use of the following objects
\begin{align}
    \bar{h}_{\mu \nu} =& h_{\mu \nu} - 2 \tau_{(\mu} m_{\nu)}\,, \label{eq:hbar}
    \\
    \bar{\Phi}_{\mu \nu} =& \Phi_{\mu \nu} - 2 \tau_{(\mu} B_{\nu)} - m_\mu m_\nu\,. \label{eq:Phibar}
\end{align}
These are local (LO and NLO) Galilean boost invariant objects that naturally show up in the expansion of the metric. 
Meanwhile, in section \ref{sec:VacSym&Bdy} we will make use of the expansion of the spatial vielbeins, so we also introduce the following notation
\begin{equation}
    \mathcal{E}^a_\mu=e^a_\mu+ \frac{1}{c^2} \pi^a_\mu + \mathcal{O}(c^{-4})\,,\qquad\mathcal{E}_a^\mu=e_a^\mu+ \frac{1}{c^2}\left(v^\mu m_\nu e^\nu_a -e^\mu_c \pi^c_\nu e^\nu_a \right) +\mathcal{O}(c^{-4})\,,
\end{equation}
where we used $T_\mu\mathcal{E}_a^\mu=0$ and $\mathcal{E}^a_\mu\mathcal{E}_b^\mu=\delta^a_b$. 

Next, we note that the PNR variables transform under both diffeomorphisms and local Lorentz boosts which act on the fields as
\begin{eqnarray}
    \delta T_\mu & = & \mathcal{L}_{\Xi} T_\mu + c^{-2}\Lambda_a\mathcal{E}^a_\mu=\mathcal{L}_{\Xi} T_\mu +c^{-2}\Lambda_\mu\,, \label{eq:GaugeTransT} \\
    \delta\Pi_{\mu\nu} & = & \mathcal{L}_{\Xi} \Pi_{\mu \nu}+ \Lambda_\mu T_\nu+\Lambda_\nu T_\mu\,, \label{eq:GaugeTransPi}
\end{eqnarray}
where $\Xi^\mu$ is a vector field generating diffeomorphisms 
and $\Lambda_\mu=\Lambda_a\mathcal{E}^a_\mu$ is the generator of local Lorentz boosts for which $T^\mu\Lambda_\mu=0$. Furthermore, we have
\begin{eqnarray}
    \delta T^\mu & = & \mathcal{L}_{\Xi} T^\mu +\Pi^{\mu\rho}\Lambda_\rho\,,  \\
    \delta\Pi^{\mu\nu} & = & \mathcal{L}_{\Xi} \Pi^{\mu \nu} + 2c^{-2}T^{(\mu}\Pi^{\nu)\rho}\Lambda_\rho\,.
\end{eqnarray}

Assuming $\Xi^\mu$ and $\Lambda_\mu$ to be analytic in $1/c^2$, we can expand the gauge parameters as 
\begin{eqnarray}
    \Xi^\mu & = & \xi^\mu+c^{-2}\zeta^\mu+\mathcal{O}(c^{-4})\,,\\
    \Lambda_\mu & = & \lambda_\mu+c^{-2}\kappa_\mu+\mathcal{O}(c^{-4})\,.
\end{eqnarray}
It then follows from \eqref{eq:GaugeTransT} and \eqref{eq:GaugeTransPi} that
\begin{eqnarray}
    \delta \tau_\mu & = & \mathcal{L}_\xi\tau_\mu\,,\label{eq:trafotau}\\
    \delta h_{\mu\nu} & = & \mathcal{L}_\xi h_{\mu\nu}+\lambda_\mu \tau_\nu+\lambda_\nu \tau_\mu\,,\label{eq:traofh}\\
    \delta m_\mu & = & \mathcal{L}_\xi m_\mu+\mathcal{L}_\zeta\tau_\mu+\lambda_\mu\,,\label{eq:traofm}\\
    \delta \Phi_{\mu\nu} & = & \mathcal{L}_\xi \Phi_{\mu\nu}+\mathcal{L}_\zeta h_{\mu\nu}+m_\mu\lambda_\nu+m_\nu\lambda_\mu+\kappa_\mu \tau_\nu+\kappa_\nu \tau_\mu\,,\label{eq:trafoPhi}
\end{eqnarray}
where $h_{\mu\nu}=\delta_{ab}e^a_\mu e^b_\nu$, $\lambda_\mu=\lambda_a e^a_\mu$ and 
\begin{equation}
    v^\mu\kappa_\mu=h^{\mu\nu}\lambda_\mu v^\rho\Phi_{\nu\rho}\,,
\end{equation}
which follows from $\delta\left(v^\mu v^\nu\Phi_{\mu\nu}\right)=0$.

Finally, it is convenient to define a covariant derivative with respect to leading order diffeomorphisms that is compatible with the Newton-Cartan metric variables, meaning
\begin{align}
    \nabla_\mu \tau_\nu = 0\,,
    \qquad
    \nabla_\mu h^{\nu \rho} =0\,.
\end{align}
These conditions still leave a lot of ambiguity in the choice of connection. However, if we also require the associated connection to only depend on the LO fields $\tau_\mu$ and $h_{\mu\nu}$ (as well as derivatives thereof), and demand that the torsion is equal to the intrinsic torsion \cite{Figueroa-OFarrill:2020gpr}, then we can show that 
\begin{align}
        \Gamma^\rho_{\mu \nu} &= \check{\Gamma}^\rho_{\mu \nu} + \alpha h^{\rho \sigma} a_\sigma \tau_\mu \tau_\nu\,,
        \\
        \check{\Gamma}^\rho_{\mu \nu}&=-v^\rho\partial_\mu\tau_\nu+\frac{1}{2}h^{\rho\sigma}\left(\partial_\mu h_{\nu\sigma}+\partial_\nu h_{\mu\sigma}-\partial_\sigma h_{\mu\nu}\right)\,, \label{eq:GammaCheck}
\end{align}
where $\alpha$ is an arbitrary real number, and where we 
\begin{equation}
    a_\mu=\mathcal{L}_v\tau_\mu\,.
\end{equation}
In deriving this we assumed that $\tau$ defines a foliation, i.e. $\tau\wedge d\tau=0$ (something that as we will see we can always assume in NRG) so that $d\tau=a\wedge \tau$. For a derivation of this see the latter half of appendix \ref{app:curvature}. For the purposes of this paper we will set $\alpha =0$ and simply make use of the $\check{\Gamma}^\rho_{\mu \nu}$-connection\footnote{Note that we use the term "connection" somewhat loosely, applying it to objects like $\check{\Gamma}^\rho_{\mu \nu}$ that still transforms under local Galilei boost.}.
We will denote the associated covariant derivative $\check{\nabla}_\mu$. We can then also define a related Riemann curvature tensor
\begin{align}
    \check{R}_{ \mu \nu \sigma}{}^\rho = - \partial_{\mu} \Check{\Gamma}^{\rho}_{\nu \sigma} + \partial_{\nu} \Check{\Gamma}^{\rho}_{\mu \sigma} -\Check{\Gamma}^{\rho}_{\mu \lambda} \Check{\Gamma}^{\lambda}_{\nu \sigma} + \Check{\Gamma}^{\rho}_{\nu \lambda} \Check{\Gamma}^{\lambda}_{\mu \sigma}\,,
\end{align}
with Ricci tensor $\check{R}_{\mu \nu} = \check{R}_{ \mu \rho \nu}{}^\rho$. 
\subsection{Curvatures for NLO fields} \label{sec:2.2}

Let us consider the gauge transformations of the NLO fields under the NLO diffeomorphisms, i.e. 
\begin{equation}\label{eq:trafosmPhi}
    \delta m_\mu=\mathcal{L}_\zeta\tau_\mu\,,\qquad \delta \Phi_{\mu\nu}=\mathcal{L}_\zeta h_{\mu\nu}\,.
\end{equation}
This is of the following general form
\begin{equation}
    \delta Q^{\alpha\beta\cdots}{}_{\mu\nu\cdots}=\mathcal{L}_\zeta S^{\alpha\beta\cdots}{}_{\mu\nu\cdots}\,,
\end{equation}
where $Q^{\alpha\beta\cdots}{}_{\mu\nu\cdots}$ is a tensor and where the tensor $S^{\alpha\beta\cdots}{}_{\mu\nu\cdots}$ is such that it is inert under the NLO diffeomorphism generated by $\zeta^\mu$, i.e.
\begin{equation}
    \delta_\zeta S^{\alpha\beta\cdots}{}_{\mu\nu\cdots}=0\,.
\end{equation}
We will define a covariant derivative $\mathcal{D}_\mu$ to be an object acting on a tensor $Q^{\alpha\beta\cdots}{}_{\mu\nu\cdots}$ such that it transforms as
\begin{equation}
    \delta_\zeta \mathcal{D}_\mu Q^{\alpha\beta\cdots}{}_{\nu\rho\cdots}=\mathcal{L}_\zeta\nabla_\mu S^{\alpha\beta\cdots}{}_{\nu\rho\cdots}\,,
\end{equation}
under NLO diffeomorphisms and where $\mathcal{D}_\mu Q^{\alpha\beta\cdots}{}_{\nu\rho\cdots}$ transforms as a tensor under LO diffeomorphisms.
For example, if we have $Q_\mu$ which transforms as $\delta_\zeta Q_\mu=\mathcal{L}_\zeta S_\mu$, then
\begin{equation}
    \mathcal{D}_\mu Q_\nu=\nabla_\mu Q_\nu-\chi^\rho_{\mu\nu}S_\rho\,,
\end{equation}
where $\chi^\rho_{\mu\nu}$ transforms under the NLO diffeomorphisms as\footnote{By $\mathcal{L}_\zeta \Gamma^\rho_{\mu\nu}$ we mean the expression one gets when applying the Lie derivative to a $(1,2)$ tensor, even though $\Gamma^\rho_{\mu\nu}$ is not an actual tensor of course.}
\begin{equation}\label{eq:trafochi}
    \delta_\zeta \chi^\rho_{\mu\nu}=\mathcal{L}_\zeta \Gamma^\rho_{\mu\nu}+\partial_\mu\partial_\nu\zeta^\rho\,.
\end{equation}
If we assign this transformation to $\chi^\rho_{\mu\nu}$ we see that we recover the desired property that 
\begin{equation}
    \delta_\zeta\mathcal{D}_\mu Q_\nu=\nabla_\mu\delta_\zeta Q_\nu-\delta_\zeta \chi^\rho_{\mu\nu} S_\rho=\mathcal{L}_\zeta \nabla_\mu S_\nu\,.
\end{equation}

So what is $\chi^\rho_{\mu\nu}$? This depends on what we take for $\Gamma^\rho_{\mu\nu}$. In the case of $\check{\Gamma}^\rho_{\mu \nu}$ of \eqref{eq:GammaCheck}, a straightforward, albeit somewhat tedious calculation, shows that \eqref{eq:trafochi} can be written as
\begin{eqnarray}
    \delta_\zeta \chi^\rho_{\mu\nu} & = & -v^\rho\check\nabla_\mu\mathcal{L}_\zeta\tau_\nu+K_{\mu\nu}h^{\rho\sigma}\mathcal{L}_\zeta\tau_\sigma+\frac{1}{2}h^{\rho\sigma}\left(\check\nabla_\mu\mathcal{L}_\zeta h_{\nu\sigma}+\check\nabla_\nu\mathcal{L}_\zeta h_{\mu\sigma}-\check\nabla_\sigma\mathcal{L}_\zeta h_{\mu\nu}\right)\nonumber\\
    &&+\frac{1}{2}v^\lambda\tau_{\mu\nu}h^{\rho\sigma}\mathcal{L}_\zeta h_{\lambda\sigma}-\frac{1}{2}v^\lambda\tau_{\mu\sigma}h^{\rho\sigma}\mathcal{L}_\zeta h_{\nu\lambda}-\frac{1}{2}v^\lambda\tau_{\nu\sigma}h^{\rho\sigma}\mathcal{L}_\zeta h_{\mu\lambda}\,,
\end{eqnarray}
where we defined
\begin{equation}
    \tau_{\mu\nu}=\partial_\mu\tau_\nu-\partial_\nu\tau_\mu\,,\qquad K_{\mu\nu}=-\frac{1}{2}\mathcal{L}_v h_{\mu\nu}\,.
\end{equation}
Using that the geometric data is inert under the $\zeta^\mu$ transformations as well as equation \eqref{eq:trafosmPhi}, we can immediately write down a choice for $\chi^\rho_{\mu\nu}$ with the desired transformation properties, namely
\begin{eqnarray}
    \chi^\rho_{\mu\nu} & = & -v^\rho\check\nabla_\mu m_\nu+K_{\mu\nu}h^{\rho\sigma}m_\sigma+\frac{1}{2}h^{\rho\sigma}\left(\check\nabla_\mu\Phi_{\nu\sigma}+\check\nabla_\nu\Phi_{\mu\sigma}-\check\nabla_\sigma\Phi_{\mu\nu}\right)\nonumber\\
    &&+\frac{1}{2}v^\lambda\tau_{\mu\nu}h^{\rho\sigma}\Phi_{\lambda\sigma}-\frac{1}{2}v^\lambda\tau_{\mu\sigma}h^{\rho\sigma}\Phi_{\nu\lambda}-\frac{1}{2}v^\lambda\tau_{\nu\sigma}h^{\rho\sigma}\Phi_{\mu\lambda}\,.
\end{eqnarray}

In order to define the curvature associated with $\mathcal{D}_\mu$ we first observe that, by definition, we have
\begin{equation}
    \mathcal{D}_\mu\mathcal{D}_\nu Q_\rho=\check\nabla_\mu\mathcal{D}_\nu Q_\rho-\chi^\sigma_{\mu\nu}\check\nabla_\sigma S_\rho-\chi^\sigma_{\mu\rho}\check\nabla_\nu S_\sigma\,,
\end{equation}
where we used that $\delta_\zeta\mathcal{D}_\mu Q_\nu=\mathcal{L}_\zeta \check\nabla_\mu S_\nu$. It follows that
\begin{equation}
    \left[\mathcal{D}_\mu\,,\mathcal{D}_\nu\right] Q_\rho=  \left[\check\nabla_\mu\,,\check\nabla_\nu\right] Q_\rho+\left(\check\nabla_\nu\chi^\sigma_{\mu\rho}-\check\nabla_\mu\chi^\sigma_{\nu\rho}\right)S_\sigma-2\chi^\sigma_{[\mu\nu]}\check\nabla_\sigma S_\rho\,.
\end{equation}
The first term, $\left[\check\nabla_\mu\,,\check\nabla_\nu\right] Q_\rho$, defines the LO curvature and torsion in terms of the Riemann and torsion tensors in the usual way. The second and third terms provide a notion of NLO torsion
\begin{equation}
2\chi^\sigma_{[\mu\nu]}=-v^\sigma\left(\check\nabla_\mu m_\nu-\check\nabla_\nu m_\mu\right)+\tau_{\mu\nu}v^\rho h^{\sigma\lambda}\Phi_{\rho\lambda}\,,    
\end{equation}
and NLO curvature, $U_{\mu\nu\rho}{}^\lambda$, with the latter defined as follows. The terms multiplying $S_\lambda$ can be written as 
\begin{equation}
    U_{\mu\nu\rho}{}^\sigma+2\check\Gamma^\kappa_{[\mu\nu]}\chi^\sigma_{\kappa\rho}\,,
\end{equation}
so that 
\begin{equation}
    \left[\mathcal{D}_\mu\,,\mathcal{D}_\nu\right] Q_\rho=\check R_{\mu\nu\rho}{}^\sigma Q_\sigma-2\check\Gamma_{[\mu\nu]}^\sigma\mathcal{D}_\sigma Q_\rho+U_{\mu\nu\rho}{}^\sigma S_\sigma-2\chi^\sigma_{[\mu\nu]}\check\nabla_\sigma S_\rho\,.
\end{equation}
We thus have
\begin{equation}\label{eq:U}
    U_{\mu\nu\rho}{}^\sigma=-\check\nabla_\mu\chi^\sigma_{\nu\rho}+\check\nabla_\nu\chi^\sigma_{\mu\rho}-2\check\Gamma^\kappa_{[\mu\nu]}\chi^\sigma_{\kappa\rho}\,.
\end{equation}
Covariance, i.e.
\begin{equation}
    \delta_\zeta\left(\left[\mathcal{D}_\mu\,,\mathcal{D}_\nu\right] Q_\rho\right)=\mathcal{L}_\zeta\left(\left[\check\nabla_\mu\,,\check\nabla_\nu\right]S_\rho\right)\,,
\end{equation}
then tells us that
\begin{equation}
    \delta_\zeta U_{\mu\nu\rho}{}^\sigma=\mathcal{L}_\zeta\check R_{\mu\nu\rho}{}^\sigma\,,
\end{equation}
as desired. This can also be verified directly by using \eqref{eq:U} and \eqref{eq:trafochi}.

Inspired by the NLO torsion $\chi^\lambda_{[\mu\nu]}$, we define the field strength
\begin{eqnarray}
    F_{\mu\nu} & = & \partial_\mu m_\nu-\partial_\nu m_\mu-a_\mu m_\nu+a_\nu m_\mu\nonumber\\
    & = & \check\nabla_\mu m_\nu-\check\nabla_\nu m_\mu-a_\mu m_\nu+a_\nu m_\mu-v^\rho m_\rho \tau_{\mu\nu}\,,\label{eq:defF}
\end{eqnarray}
which transforms as
\begin{equation}\label{eq:trafohhF}
   h^{\mu\rho}h^{\nu\sigma} \delta_\zeta F_{\mu\nu}=h^{\mu\rho}h^{\nu\sigma}\mathcal{L}_\zeta \left(P\tau\right)_{\mu\nu}\,,
\end{equation}
where we defined
\begin{equation}
    \left(P\tau\right)_{\mu\nu}=P^\kappa_\mu P^\lambda_\nu\tau_{\kappa\lambda}\,,
\end{equation}
with
\begin{equation}
    P^\kappa_\mu=\delta^\kappa_\mu+v^\kappa \tau_\mu\,,
\end{equation}
a projector. We thus see that the spatial projection of $F_{\mu\nu}$ is gauge invariant for a TTNC LO geometry\footnote{We remind the reader that TTNC geometry stands for twistless torsional Newton-Cartan geometry \cite{Christensen:2013lma} and is referring to a geometry that fulfills \eqref{eq:TTNC}.}. In appendix \ref{app:ansatz} we define a gauge invariant field strength for $\Phi_{\mu\nu}$ for a LO geometry given by our vacuum solution. 

\subsection{Lagrangian of non-relativistic gravity} \label{sec:2.3}

Using what we have learned it is now very straightforward to build actions that are invariant under the LO and NLO diffeomorphisms. For example, the Lagrangian density
\begin{equation}\label{eq:NLOLagrangian_part}
    e\left(h^{\mu\rho}U_{\mu\lambda\rho}{}^\lambda+\left(-h^{\mu\alpha}h^{\nu\beta}\Phi_{\alpha\beta}+2h^{\alpha(\mu}m_\alpha v^{\nu)}\right)\check R_{\mu\nu}+I h^{\mu\nu}\check R_{\mu\nu}\right)\,,
\end{equation}
where $e=\text{det}\,(\tau_\mu\,, e^a_\mu)$ and where we introduced
\begin{equation}
    I=-v^\rho m_\rho+\frac{1}{2}h^{\rho\sigma}\Phi_{\rho\sigma}\,,
\end{equation}
transforms into a total derivative under NLO diffeomorphisms. To see this it is useful to note that 
\begin{equation}
    \delta_\zeta\left(-h^{\mu\alpha}h^{\nu\beta}\Phi_{\alpha\beta}+2h^{\alpha(\mu}m_\alpha v^{\nu)}\right)=\mathcal{L}_\zeta h^{\mu\nu}\,,
\end{equation}
which follows from the fact that the left-hand side is the NLO term in the expansion of $\Pi^{\mu\nu}$. Additionally,
\begin{equation}
\delta_\zeta I=e^{-1}\partial_\rho(e\zeta^\rho)\,,
\end{equation}
which follows from the fact that $\sqrt{-g}=c e\left(1+c^{-2}I+\mathcal{O}(c^{-4})\right)$. Furthermore, the Lagrangian in \eqref{eq:NLOLagrangian_part} is obviously a tensor density with respect to LO diffeomorphisms so this can be used as a building block to construct Lagrangians that are invariant under the full set of transformations in \eqref{eq:trafotau}--\eqref{eq:trafoPhi}.

Since \eqref{eq:NLOLagrangian_part} already transform correctly under LO and NLO diffeomorphisms all we need to do is to ensure that we add appropriate terms to \eqref{eq:NLOLagrangian_part} to make the total Lagrangian invariant under LO and NLO Galilean boosts (with parameters $\lambda_\mu$ and $\kappa_\mu$) whilst ensuring that the extra terms transform correctly under LO and NLO diffeomorphisms. This process leads to an action that can also be obtained from the $1/c^2$ expansion of the EH Lagrangian (as was also done in \cite{Hansen:2020pqs}). We will review this procedure now.

In PNR variables the EH action is \cite{Hansen:2020pqs}
\begin{equation}
    S=\frac{c^6}{16\pi G}\int d^{D}x E\left(\frac{1}{4}\Pi^{\mu\nu}\Pi^{\rho\sigma}T_{\mu\rho}T_{\nu\sigma}+c^{-2}\left(\Pi^{\mu\nu}\overset{(C)}{R}_{\mu\nu}-2\Lambda\right)-c^{-4} T^\mu T^\nu\overset{(C)}{R}_{\mu\nu}\right)\,, \label{eq:PNRaction}
\end{equation}
where we have included a cosmological constant $\Lambda$ that we will take to be independent of $c$, and where we defined $E = \det(T_\mu, \mathcal{E}^a_\mu)$ as well as
\begin{equation}
    T_{\mu\nu}=\partial_\mu T_\nu-\partial_\nu T_\mu\,.
\end{equation}
The $(C)$ above an object means that it is calculated with respect to the PNR connection $C^\rho_{\mu\nu}$ defined in \eqref{C-conn}. The expansion of the $C^\rho_{\mu\nu}$ connection is given by
\begin{equation}
    C^\rho_{\mu\nu}=\check\Gamma^\rho_{\mu\nu}+c^{-2}\chi^\rho_{\mu\nu}+\mathcal{O}(c^{-4})\,.
\end{equation}
While the associated Riemann tensor expands as
\begin{equation}
 \overset{(C)}{R}_{\mu\nu\rho}{}^\sigma=\check R_{\mu\nu\rho}{}^\sigma+c^{-2}U_{\mu\nu\rho}{}^\sigma+\mathcal{O}(c^{-4})\,.   
\end{equation}
The Lagrangian \eqref{eq:NLOLagrangian_part} is thus the NLO term in the expansion of $E\Pi^{\mu\nu}\overset{(C)}{R}_{\mu\nu}$.

The NLO term in the expansion of \eqref{eq:PNRaction} is given by
\begin{equation}
    S_{\text{NLO}}=\frac{c^4}{16\pi G}\int d^{D}x \left[\frac{1}{4}E\Pi^{\mu\nu}\Pi^{\rho\sigma}T_{\mu\rho}T_{\nu\sigma}\Big|_{\text{NLO}}+E\left(\Pi^{\mu\nu}\overset{(C)}{R}_{\mu\nu}-2\Lambda\right)\Big|_{\text{LO}}\right]\,,
\end{equation}
where 
\begin{eqnarray}
E\left(\Pi^{\mu\nu}\overset{(C)}{R}_{\mu\nu}-2\Lambda\right)\Big|_{\text{LO}} & = & e\left(h^{\mu\nu}\check R_{\mu\nu}-2\Lambda\right)\,,\\
    \frac{1}{4}E\Pi^{\mu\nu}\Pi^{\rho\sigma}T_{\mu\rho}T_{\nu\sigma}\Big|_{\text{NLO}} & = & e\left(\frac{1}{2}h^{\mu\rho}h^{\nu\sigma}F_{\mu\nu}\tau_{\rho\sigma}+\frac{1}{4}I h^{\mu\rho}h^{\nu\sigma}\tau_{\mu\nu}\tau_{\rho\sigma}-\frac{1}{2}h^{\kappa\lambda}h^{\mu\rho}h^{\nu\sigma}\tau_{\mu\kappa}\tau_{\rho\sigma}\Phi_{\lambda\nu}\right)\,.\nonumber\\
    &&\,.\label{eq:NLOpartofPiPITT}
\end{eqnarray}
We note that the component $v^\mu m_\mu$ appears algebraically in the NLO action and enforces the TTNC condition. The NNLO term in the expansion of \eqref{eq:NLOLagrangian_part} is given by
\bea{}
    S_{\text{NNLO}}=\frac{c^2}{16\pi G}\int d^{D}x \Big[\frac{1}{4}E\Pi^{\mu\nu}\Pi^{\rho\sigma}T_{\mu\rho}T_{\nu\sigma}\Big|_{\text{NNLO}}+E(\Pi^{\mu\nu}\overset{(C)}{R}_{\mu\nu}-2\Lambda)\Big|_{\text{NLO}}\nonumber\\ -ET^\mu T^\nu\overset{(C)}{R}_{\mu\nu}\Big|_{\text{LO}}\Big]\,,
\eea
where 
\begin{eqnarray}
    \frac{1}{4}E\Pi^{\mu\nu}\Pi^{\rho\sigma}T_{\mu\rho}T_{\nu\sigma}\Big|_{\text{NNLO}} & = & e\left(\frac{1}{4}h^{\mu\nu}h^{\rho\sigma}F_{\mu\rho}F_{\nu\sigma}+h^{\mu\nu}h^{\rho\sigma}\tau_{\mu\rho}X_{\nu\sigma}\right)\,,\label{eq:PiPiTT}\\
    E\left(\Pi^{\mu\nu}\overset{(C)}{R}_{\mu\nu}-2\Lambda\right)\Big|_{\text{NLO}} & = & e\left(h^{\mu\rho}U_{\mu\lambda\rho}{}^\lambda+I\left( h^{\mu\nu}\check R_{\mu\nu}-2\Lambda\right)+\right.\nonumber\\
    &&\left.\left(-h^{\mu\alpha}h^{\nu\beta}\Phi_{\alpha\beta}+2h^{\alpha(\mu}m_\alpha v^{\nu)}\right)\check R_{\mu\nu}\right)\,,\label{eq:bareNLOterm}\\
    ET^\mu T^\nu\overset{(C)}{R}_{\mu\nu}\Big|_{\text{LO}} & = & ev^\mu v^\nu\check R_{\mu\nu}\,,
\end{eqnarray}
where equation \eqref{eq:bareNLOterm} is precisely \eqref{eq:NLOLagrangian_part} with a cosmological constant added. We also see that the field $\phi=-v^\mu m_\mu$ once again appears algebraically in the action and enforces the condition that 
\begin{equation}\label{eq:Phi-EOM}
    h^{\mu\nu}\check R_{\mu\nu}=2\Lambda\,.
\end{equation}
The field $\phi$ is the field that encodes the Newtonian potential in the case of absolute time ($d \tau =0$).

Finally, the antisymmetric field $X_{\nu\sigma}$ in \eqref{eq:PiPiTT} has been left unspecified because the total variation of the part in the NNLO action containing $X_{\nu\sigma}$ is 
\begin{eqnarray}
    \delta S_X & = & \frac{c^2}{16\pi G}\int d^{D}x\left( e\left[-v^\rho h^{\mu\nu}h^{\rho\sigma}\tau_{\mu\rho}X_{\nu\sigma}-2e^{-1}\partial_\mu\left(e h^{\mu\nu}h^{\rho\sigma}X_{\nu\sigma}\right)\right]\delta\tau_\rho\right.\nonumber\\
    &&\left.e\left[\frac{1}{2}h^{\alpha\beta}h^{\mu\nu}h^{\rho\sigma}\tau_{\mu\rho} X_{\nu\sigma}+2h^{\mu\alpha}h^{\nu\beta}h^{\rho\sigma}\tau_{\rho(\mu}X_{\nu)\sigma}\right]\delta h_{\alpha\beta}\right.\nonumber\\
    &&\left.e h^{\mu\nu}h^{\rho\sigma}\tau_{\mu\rho}\delta X_{\nu\sigma}\right)\,,
\end{eqnarray}
where we emphasise that $X_{\nu\sigma}$ is not a fundamental variable but something that depends on various fields appearing in the $1/c^2$ expansion of the metric. With the exception of one term, all terms in the variation of $S_X$ are proportional to $(P\tau)_{\mu\nu}$. In $X_{\nu\sigma}$ there will be NNLO fields that enforce the equations of motion of the LO action\footnote{The NNLO fields $B_\mu$ and $\psi_{\mu\nu}$ appear in the NNLO action precisely like the NLO fields $m_\mu$ and $\Phi_{\mu\nu}$ appear in the NLO action. Specifically, if we take \eqref{eq:NLOpartofPiPITT} and replace $m_\mu$ with $B_\mu$ and $\Phi_{\mu\nu}$ with $\psi_{\mu\nu}$ we obtain the part of the NNLO action that contains the NNLO fields. The reason behind this is that in the NNLO action a NNLO field can only come multiplied by LO fields. So for the purpose of just computing these terms we can set the NLO fields equal to zero. If we do that the NNLO fields can be viewed as the NLO fields of an expansion with the parameter $c^{-4}$ and so the NNLO fields in the NNLO action appear just like the NLO fields in the NLO action (when we expand in $c^{-2}$).} \cite{Hansen:2019svu} and so these will tell us that $\tau_\mu$ obeys the TTNC condition. Hence after performing the variation with respect to the NNLO fields we can drop terms proportional to $(P\tau)_{\mu\nu}$. This leaves us with 
\begin{equation}
    \delta S_X \Big\vert_{\text{TTNC}}= -\frac{c^2}{8\pi G}\int d^{D}x \partial_\mu\left(e h^{\mu\nu}h^{\rho\sigma}X_{\nu\sigma}\right)\delta\tau_\rho\,.
\end{equation}
This will lead to an equation of motion for the NNLO fields appearing inside $X_{\nu\sigma}$ and therefore, if we are not interested in NNLO fields, we can ignore this. More specifically, if we vary $\tau_\rho$ in the direction of $\tau_\rho$, i.e. if we take $\delta\tau_\rho=\Omega\tau_\rho$ with $\Omega$ arbitrary then the variation of $S_X$ upon using TTNC vanishes. This means that for the purpose of computing the EOM that do not contain NNLO fields we should include the variation $\delta\tau_\rho=\Omega\tau_\rho$, but not $P^\kappa_\rho\delta\tau_\kappa$.

We will from now on treat $X_{\nu\sigma}$ as in an independent field and call it $\zeta_{\nu\sigma}$ which will play the role of a Lagrange multiplier enforcing the TTNC condition. This procedure gives us what we call the NRG Lagrangian:
\begin{eqnarray}\label{eq:action_NRG}
\hspace{-1.5cm}\mathcal{L}_\mathrm{NRG} & \equiv & \left.\mathcal{L}_{\text{NNLO}}\right|_{\tau\wedge d\tau=0}+
\frac{e}{16\pi G_N}\zeta_{\rho\sigma}h^{\mu\rho}h^{\nu\sigma}(\partial_\mu\tau_\nu-\partial_\nu\tau_\mu)\nonumber\\
\hspace{-1.5cm}& = & \frac{e}{16\pi G_N}\Bigg[h^{\mu\rho}h^{\nu\sigma}K_{\mu\nu}K_{\rho\sigma}-\left(h^{\mu\nu}K_{\mu\nu}\right)^2
-2m_\nu\left(h^{\mu\rho}h^{\nu\sigma}-h^{\mu\nu}h^{\rho\sigma}\right)\check{\nabla}_\mu K_{\rho\sigma}
\nonumber\\
\hspace{-1.5cm}&&+\phi h^{\mu\nu}\check R_{\mu\nu}
+\frac{1}{4}h^{\mu\rho}h^{\nu\sigma}F_{\mu\nu}F_{\rho\sigma}+\frac{1}{2}\zeta_{\rho\sigma}h^{\mu\rho}h^{\nu\sigma}(\partial_\mu\tau_\nu-\partial_\nu\tau_\mu)\nonumber\\
\hspace{-1.5cm}&&
-\Phi_{\rho\sigma}h^{\mu\rho}h^{\nu\sigma}\left(\check R_{\mu\nu}-\check{\nabla}_\mu a_\nu - a_\mu a_\nu
-\frac{1}{2}h_{\mu\nu}h^{\kappa\lambda}\check R_{\kappa\lambda}
+h_{\mu\nu} e^{-1} \partial_{\kappa}\left(e h^{\kappa\lambda} a_\lambda\right)\right)\Bigg]\,.
\end{eqnarray}
In writing this we  used the explicit form for $U_{\mu\lambda\rho}{}^\lambda$ as well as the identity
\begin{equation}
v^\mu v^\nu\check R_{\mu\nu}=\left(h^{\mu\nu}K_{\mu\nu}\right)^2-h^{\mu\rho}h^{\nu\sigma}K_{\mu\nu}K_{\rho\sigma}+\check\nabla_\rho\left(v^\rho h^{\mu\nu}K_{\mu\nu}\right)\,.
\end{equation}
We can add a cosmological constant term by adding to NRG the following Lagrangian\footnote{We assumed here that in the $1/c^2$ expansion $\Lambda$ was treated as independent of $c$, so that $\Lambda$ appears in the NLO theory for the first time. Alternatively we could have assumed that $\Lambda=\mathcal{O}(c^{-2})$ so that it appears in the NNLO theory for the first time as $-\frac{\Lambda}{8\pi G}e$. This way of doing things leads to Newton--Hooke spacetimes \cite{Grosvenor:2017dfs,Hansen:2020pqs} which have absolute time.}
\begin{equation} \label{eq:LambdaAction}
    \mathcal{L}_{\text{NRG}-\Lambda}=-\frac{\Lambda}{8\pi G}e I\,.
\end{equation}

In appendix \ref{app:EOM} we show that the equations of motion of NRG with a cosmological constant can be written in the following form:
\begin{eqnarray}
e^{-1}\partial_\mu\left(eh^{\mu\nu}a_\nu\right) & = & -\frac{2}{D-2}\Lambda\,,\label{eq:diva}\\
    e^{-1}\partial_\mu\left(e\tilde X^\mu\right)+v^\mu v^\nu\check R_{\mu\nu}+\frac{1}{4} h^{\mu\nu}h^{\rho\sigma}F_{\mu\rho}F_{\nu\sigma}
    & = & -\frac{2}{D-2}\Lambda I\,,\label{eq:Pois}\\
-2h^{\nu\sigma}v^\alpha\check R_{\nu\alpha}+\left(\check\nabla_\rho+2a_\rho\right)h^{\rho\mu}h^{\sigma\nu}F_{\mu\nu} & = & 0\,,\label{eq:PvNLOEOM}\\
h^{\mu\kappa} h^{\nu\lambda}\left(\check R_{\mu\nu}-\check\nabla_\mu a_{\nu}-a_\mu a_\nu\right) & = & \frac{2}{D-2}\Lambda h^{\kappa\lambda}\,,\label{eq:NLO-EOMhh}
\end{eqnarray}
\begin{eqnarray}
    &&h^{\mu\kappa} h^{\nu\lambda}\left(m_\mu v^\rho\left(\check R_{(\nu\rho)}-\check\nabla_{(\nu}a_{\rho)}\right)+m_\nu v^\rho\left(\check R_{(\mu\rho)}-\check\nabla_{(\mu}a_{\rho)}\right)-  U_{\alpha(\mu\nu)}{}^\alpha+\chi^\rho_{\mu\nu}a_\rho\right. \label{eq:NNLO-EOMhh}\\
    &&\left.+v^\rho\left(\check\nabla_{(\mu}+2a_{(\mu}\right)F_{\nu)\rho}+K K_{\mu\nu}-v^\rho\check\nabla_\rho K_{\mu\nu}+\frac{1}{2}h^{\rho\sigma}F_{\mu\rho}F_{\nu\sigma}\right)=\frac{2}{D-2}\Lambda h^{\mu\kappa} h^{\nu\lambda} \Phi_{\mu\nu}\,,\nonumber
\end{eqnarray}
where $K = h^{\mu \nu} K_{\mu \nu}$ and where in \eqref{eq:Pois} we defined
\begin{equation}
    \tilde X^\mu =  h^{\mu\sigma}v^\rho\left(\partial_\rho m_\sigma-\partial_\sigma m_\rho\right)+v^\mu h^{\rho\sigma}a_\rho m_\sigma-\frac{1}{2}h^{\mu\kappa}h^{\nu\lambda}\left(a_\nu\Phi_{\lambda\kappa}+a_\lambda\Phi_{\nu\kappa}-a_\kappa\Phi_{\nu\lambda}\right)\,.
\end{equation}
Note that the EOM, particularly equation (\ref{eq:NNLO-EOMhh}), have shrunk significantly in size to those given in \cite{Hansen:2020pqs} (see \eqref{eq:NRG_EOM4} for the original equation). This was mainly due to the realisation that for specific EOM, particularly for the variation with respect to $h_{\mu \nu}$, the EOM of the other fields can be used to get rid of a significant amount of terms. The introduction of the NLO curvature leads to further simplification and it will also prove a useful object later on when we construct the non-relativistic AdS brane solution.

\section{Vacuum, symmetries and boundary} \label{sec:VacSym&Bdy}
In this section we set up the basic elements needed to begin constructing a non-relativistic fluid/gravity correspondence. First we consider the vacuum solution, namely non-relativistic AdS, and work out its isometries which correspond to an infinite lift of the finite dimensional Galilean conformal algebra (GCA). Though, as we will see, this infinite lift is an emergent symmetry stemming from the order at which we have truncated the expansion. Finally, we construct a conformal boundary to a class of non-relativistic spacetimes that are asymptotically locally non-relativistic AdS and then show that we get a conformal class of type II Newton-Cartan geometries on this boundary.

\subsection{Vacuum} \label{sec:3.1}

The vacuum solution is non-relativistic AdS which is given by 
\begin{align}
    &\tau_\mu dx^\mu = \frac{R}{l} dt\,,
    \quad
    h_{\mu \nu} dx^\mu dx^\nu=\frac{l^2}{R^2} dR^2 + \frac{R^2}{l^2} d \vec{x}^2\,, \nonumber
    \\
    &m_\mu dx^\mu = 0\,,
     \qquad
     \Phi_{\mu \nu} dx^\mu dx^\nu= 0\,, \label{eq:NRAdS}
\end{align} 
where $d\vec{x}=(dx^1,\cdots,dx^d)$ with $d=D-2$ being the total number of boundary spatial dimensions.

If we define $r=l^2/R$ the vacuum is given by
\begin{equation}
    \tau_\mu = \frac{l}{r}\delta^t_\mu\,,\qquad h_{\mu\nu}=\frac{l^2}{r^2}\left(\delta^r_\mu\delta^r_\nu+\delta^i_\mu\delta^i_\nu\right)\,. \label{eq:NRAdS2}
\end{equation}
We then have for the inverse objects
\begin{equation}
    v^\mu = -\frac{r}{l}\delta^\mu_t\,,\qquad h^{\mu\nu}=\frac{r^2}{l^2}\left(\delta^\mu_r\delta^\nu_r+\delta^\mu_i\delta^\nu_i\right)\,, \label{eq:NRAdS3}
\end{equation}
where we use $i,j,k=1,\cdots,d$ to denote the indices of the boundary spatial dimensions. It follows that in these coordinates we have
\begin{eqnarray}
    a_\mu & = & -r^{-1}\delta_\mu^r\,,\\
    \check\Gamma^\rho_{\mu\nu} & = & v^\rho\tau_\mu a_\nu+a_\mu\delta^\rho_\nu+a_\nu\delta^\rho_\mu-h^{\rho\sigma}a_\sigma h_{\mu\nu}\,.
\end{eqnarray}
With this we can show that
\begin{eqnarray}
    \check R_{\mu\nu\rho}{}^\sigma & = & -l^{-2}h^{\sigma\lambda}\left(h_{\mu\rho}h_{\nu\lambda}-h_{\mu\lambda}h_{\nu\rho}\right)\,,\label{eq:vacRiem}\\
    K_{\mu\nu} & = & 0\,.\label{eq:vacK}
\end{eqnarray}
This will be useful in later calculations.

\subsection{Symmetries} \label{sec:3.2}

We want to study the isometries of this non-relativistic AdS spacetime. These are the allowed diffeomorphisms, parameterised by $\xi^\mu$, for which 
\begin{subequations}
\begin{align}
    0&= \delta \tau_\mu = \mathcal{L}_\xi \tau_\mu, \label{eq:tauinv}
    \\
    0 &= \delta h_{\mu \nu} =\mathcal{L}_\xi h_{\mu \nu} + 2 \lambda_{(\mu} \tau_{\nu)}, \label{eq:hinv}
    \\
    0& = \delta m_\mu = \mathcal{L}_{\xi} m_\mu + \mathcal{L}_{\zeta} \tau_\mu + \lambda_\mu, \label{eq:minv}
    \\
    0& = \delta \Phi_{\mu \nu} = \mathcal{L}_{\xi} \Phi_{\mu \nu} + \mathcal{L}_{\zeta} h_{\mu \nu} + 2 \lambda_{(\mu} m_{\nu)} + 2 \kappa_{(\mu} \tau_{\nu)} \,.
    \label{eq:phiinv}
\end{align}
\end{subequations}
Solving equations (\ref{eq:tauinv}) and (\ref{eq:hinv}) we get
\begin{align}
    \xi^t &= F(t),
    \qquad 
    \xi^R = -  R \dot F(t)\,.
    \qquad
    \xi^i = A^i(t) + A^{ij}(t) x^j + \dot F(t) x^i, \label{eq:LOxi}
    \\
   \lambda_i &=- \frac{R}{l} \big(\dot A^i(t) + \dot A^{ij}(t) x^j + \ddot F(t) x^i \big)\,,
   \qquad
    \lambda_R = \frac{l^3}{R^2} \ddot F(t)\,,
\end{align}
where $F(t),\, A^i(t)$ and $A^{ij}(t)$ are arbitrary functions of time and $A^{(ij)} = 0$. One might naively think that this is the final solution for $\xi^{\mu}$ since equation \eqref{eq:minv} and \eqref{eq:phiinv}
do not depend explicitly on $\xi^\mu$ ($m_\mu=0$ and $\Phi_{\mu \nu}=0$). However, through $\lambda_\mu$ the functions in $\xi^\mu$ get constrained even further by equation \eqref{eq:minv} and \eqref{eq:phiinv}.

Specifically, solving equation (\ref{eq:minv}) and \eqref{eq:phiinv} we find 
\begin{align}
    \dddot F(t) = 0\,, \qquad \dot A^{ij} =0\,,
\end{align}
as well as expressions for $\zeta^\mu$ and $\kappa_\mu$ such that in total we have
\begin{subequations} \label{eq:GCAlift}
\begin{align}
    \xi^t &= F_0 + F_1 t + F_2 t^2\,,
    \\
    \xi^R &= - R (F_1 + 2F_2 t)\,,
    \\
    \xi^i &= A^i(t) + A^{ij} x^j +  x^i (F_1 + 2F_2 t)\,,
    \\
    \zeta^t &= \dot A_i(t) x^i + \bigg(  x^2 + \frac{l^4}{R^2}\bigg) F_2 + G(t)\,, \label{eq:NLOzetat}
    \\
    \zeta^R & = - R \bigg(\ddot A_i(t) x^i  + \dot G(t) \bigg)\,, \label{eq:NLOzetaR} 
    \\
    \zeta^i &= - \frac{1}{2} \frac{l^4}{R^2} \ddot A^i(t) + x^i x_k \ddot A^k(t) - \frac{1}{2} x^2 \ddot{A}^i(t) + \dot G(t) x^i + B_{ik}(t)x^k+B^i(t)\,, \label{eq:NLOzetai}
    \\
    \lambda_i &=  - \frac{R}{l} \big(\dot A^i(t) + 2 F_2 x^i \big)\,, \qquad \lambda_R = \frac{2l^3}{R^2}  F_2\,,
    \\
    \kappa_i &= - \frac{R}{l} \partial_t \zeta^i\,,
    \qquad \kappa_R = - \frac{l^3}{R^3} \partial_t \zeta^R\,, 
\end{align}
\end{subequations}
where we have introduced the notation $F(t) = F_0 + F_1t + F_2 t^2$, and where $G(t)$, $B_{ij}(t)$ and $B^i(t)$ are arbitrary functions of time with $B^{(ij)}=0$. 

Notably, there is still an arbitrary function of time, $A^i(t)$, in $\xi^i$. So the Killing symmetries form a infinite lift of the finite dimensional Galilean conformal algebra (GCA). This is related to the fact that we have chosen to truncate the $1/c^2$-expansion at NNLO. If we had truncated the expansion at one order higher we would get two additional fields from the expansion of the metric, $B_\mu$ and $\Psi_{\mu \nu}$, along with two additional equations to constrain the isometries of non-relativistic AdS
\begin{align}
    &0=\delta B_\mu = \mathcal{L}_{\gamma} \tau_\mu + \mathcal{L}_{\zeta} m_\mu +\mathcal{L}_{\xi} B_\mu+ \kappa_\mu \,, 
    \\ 
    &0=\delta \Psi_{\mu \nu} = \mathcal{L}_{\gamma} h_{\mu \nu}+ \mathcal{L}_{\zeta} \Phi_{\mu \nu} +\mathcal{L}_{\xi} \Psi_\mu + 2 \theta_{(\mu}\tau_{\nu)} + 2 \kappa_{(\mu}m_{\nu)}+ 2 \lambda_{(\mu}B_{\nu)}\,,
\end{align}
where $\gamma^\mu$ and $\theta_\mu$ are the NNLO diffeomorphism and boost parameters, respectively. Through $\kappa_\mu$ these equation would constrain $A^i(t)$. More specifically, we would find that\footnote{With the addition of this constraint the Killing vectors form the finite dimensional GCA, as one might expect.} $\dddot A^i (t) = 0$. Additionally, it would also constrain the subleading diffeomorphism parameters further with $\dddot G(t) = 0$ and $\dot B^{ij} = 0$. Of course, going to one order higher we also introduce new arbitrary functions of time in the NNLO diffeomorphism parameters. 

At this point, one might notice that a pattern starts to emerge. Truncating the expansion at a given order, the most subleading diffeomorphism parameters will always contain a set of time dependent functions unique to that order (like $\{F(t), A^i(t), A^{ij}(t)\}$ in \eqref{eq:LOxi} for truncation at LO or $\{G(t), B^i(t) , B^{ij}(t) \}$  in \eqref{eq:NLOzetat}-\eqref{eq:NLOzetai} 
for truncation at NLO). However, if one then extends the expansion by two more orders, all these functions will be constrained to be finite polynomials in time. Yet, this extension of the expansion simultaneously introduces new arbitrary functions of time in diffeomorphism parameters at the now most and next-to-most subleading orders. These can be seen as emergent symmetries.

\subsection{Boundary} \label{sec:3.3}
Next, we want to discuss the definition of a boundary for non-relativistic spacetimes that are asymptotically 
 locally non-relativistic AdS. So we start by considering a $(d+2)$-dimensional manifold $\mathcal{M}$ endowed with a Newton-Cartan metric structure ($\tau,h$) or the equivalent Newton-Cartan tetrad ($\tau$,$e^a$). 
For non-relativistic spacetimes that are asymptotically locally AdS we are always free to choose a gauge such that 
\begin{align}
    &e^{d+1} = \frac{l}{r} dr\,,
    \qquad
    \tau_r = 0\,,
    \qquad
    e^a_r =0\,, \label{eq:GaugeChoice1}
    \\
    &m_r =0\,, \qquad \pi^{a}_r = 0\,, \qquad \pi^{d+1}_\mu = 0\,, \label{eq:GaugeChoice2}
\end{align}
where the radial coordinate $r$ is chosen such that the boundary lies at $r=0$ and where $\tau_M, e^a_N \sim \mathcal{O}(r^{-1})$ for $N,M=(t,i)$. It follows from this that for the inverse tetrad
\begin{align}
    e_{d+1} = \frac{r}{l} \partial_r\,, \qquad v^r =0\,, \qquad e_a^r = 0\,.
\end{align}

We can then define a conformal boundary for the asymptotically locally non-relativistic AdS spacetimes in the usual sense of constructing a compactified manifold $\bar{\mathcal{M}}$ with a boundary $\partial \bar{\mathcal{M}}$ and whose interior is $\mathcal{M}$. The two manifolds ${\mathcal{M}}$ and $\bar{\mathcal{M}}$ will be linked by the defining function $\Omega$ which satisfies $\Omega>0$ in the interior of $\bar{\mathcal{M}}$ while $\Omega =0$ and $d\Omega \neq 0$ on $\partial \bar{\mathcal{M}}$. The non-relativistic vielbein on $\bar{\mathcal{M}}$ are then given by 
\begin{align}
    \bar \tau = \Omega \tau\,, \qquad \bar e^a = \Omega e^a\,, \qquad \bar e^{d+1} = \Omega e^{d+1}\,.
\end{align}
where $\bar{e}^{d+1} \propto d \Omega$ is the normal 1-form to the boundary. 
We can then define the boundary tetrad by
\begin{align}
    &\tau_\partial := \bar \tau|_{\Omega=0} \,, \qquad  e^a_\partial = \bar e^a|_{\Omega =0}\,,
    \\
    &m_\partial := \bar m|_{\Omega=0} \,, \qquad  \pi^a_\partial = \bar \pi^a|_{\Omega =0}\,,
\end{align}
A natural choice for the defining function is $\Omega = \frac{r}{l}$\,\footnote{This is of course not a unique choice, one is allowed to use any defining function $\Omega'$ for which $\Omega'= e^{f(x)}\Omega$.}, which is the choice we are going to use in the rest of this section.

In order to better understand the properties of the boundary geometry we will work out the gauge transformations of the boundary vielbein. For this we must first compute the residual gauge transformations of the choice made in \eqref{eq:GaugeChoice1} and \eqref{eq:GaugeChoice2}. For the leading order vielbein the conditions become
\begin{align}
    0 &= \delta \tau_r = \mathcal{L}_\xi \tau_r,
    \qquad
    0 = \delta e^a_r = \mathcal{L}_{\xi} e_r^a + \lambda^a{}_{d+1} e^{d+1}_r
    \\
    0 &= \delta e^{d+1}_\mu = \mathcal{L}_{\xi} e_\mu^{d+1} + \lambda^{d+1}{}_{a} e^{a}_\mu + \lambda^{d+1} \tau_\mu
\end{align}
This is solved by
\begin{align}
    \xi^r =& \frac{r}{l} \omega(t,x^i)\,, \qquad \xi^M = \chi^\mu (t,x^i) - \int_0^r dr' \frac{l}{r'} h^{M N} \partial_N \omega(t,x^i)\,,
    \\
    \lambda^{d+1}{}_c =& - e^{M}_c \partial_M \omega(t,x^i)\,, \qquad \lambda^{d+1} = v^M \partial_M \omega(t,x^i)\,,
\end{align}
where we see that $\chi^\mu(t,x^i)$ looks like a boundary diffeomorphisms and $\omega(t,x^i)$ correspond to a Weyl transformation. In other words, this is the leading order term in the $1/c^2$-expansion of a Penrose-Brown-Henneaux (PBH) transformation \cite{Penrose:1986ca,Brown:1986nw}.

At next-to-leading order the choice is preserved by gauge transformations for which
\begin{align}
    0 &= \delta m_r = \mathcal{L}_\xi m_r + \mathcal{L}_\zeta \tau_r + \lambda_{d+1} e^{d+1}_r \,, 
    \qquad 
    0= \delta \pi^{a}_r = \mathcal{L}_\xi \pi_r^a + \mathcal{L}_\zeta e^a_r + \kappa^a{}_{d+1} e^{d+1}_r \,,  \nonumber
    \\ 
    0&= \delta \pi^{d+1}_\mu= \mathcal{L}_\xi \pi_\mu^{d+1} + \mathcal{L}_\zeta e^{d+1}_\mu+\lambda^{d+1} m_\mu+\kappa^{d+1} \tau_\mu + \lambda^{d+1}{}_{a} \pi^{a}_\mu+\kappa^{d+1}{}_{a} e^{a}_\mu \,. 
\end{align}
From this we find that 
\begin{align}
    \zeta^r &= \frac{r}{l} \sigma(t,x^i)\,, \nonumber
    \\
    \zeta^M &=\eta^M (t,x^i)-  \int_0^r dr' \frac{l}{r'} \left[ h^{M N} \partial_N \sigma(t,x^i)\right. \nonumber\\
    &\left.+ \left(h^{M P} h^{N Q} \Phi_{P Q} - 2 m_P h^{P(M} v^{N)} + v^M v^N \right) \partial_N \omega(t,x^i) \right] \,, \nonumber
    \\
    \kappa^{d+1} &=v^M \partial_M \sigma(t,x^i) + v^M v^N m_M \partial_N \omega(t,x^i) - v^M \pi_M^a e^N_a \partial_N \omega(t,x^i) \,, \nonumber
    \\
    \kappa^{d+1}{}_a &= e^M_ a \partial_M \sigma(t,x^i) + e^M_a (m_M v^N - \pi^a_M e^N_a) \partial_N \omega(t,x^i) \,,
\end{align}
where $\eta(t,x^i)$ acts like a subleading boundary diffeomorphism and $\sigma(t,x^i)$ as a subleading Weyl transformation. Again this is consistent with the $1/c^2$-expansion of a Penrose-Brown-Henneaux (PBH) transformation.

Finally, we use that $\delta \tau_\partial = \big(\frac{1}{\Omega} \delta \tau\big)|_{\Omega=0}$ and equivalent statements for the other vielbein to see that the boundary gauge transformations act as follows on the boundary vielbein
\begin{align}
    \delta \tau &=  \omega \tau + \mathcal{L}_\chi \tau\,, \label{eq:Bdy_trans1}
    \\
    \delta e^a &=  \omega e^a + \mathcal{L}_\chi e^a + \lambda^a \tau + \lambda^a{}_b e^b\,,
    \\
    \delta m &= \omega m 
    + \sigma \tau + \mathcal{L}_\chi m + \mathcal{L}_\eta \tau + \lambda_a e^a\,,
    \\
    \delta \pi^a &=  \omega \pi^a 
    + \sigma e^a + \mathcal{L}_\chi \pi^a + \mathcal{L}_\eta e^a + \lambda^a m + \kappa^a \tau + \lambda^a{}_b \pi^b + \kappa^a{}_b e^b\, \label{eq:Bdy_trans2}
\end{align}
where we have omitted all subscript $\partial$ to reduce notational clutter since everything is evaluated at the boundary ($\Omega =0$) in the above statement. For similar reasons we have also rescaled the parameters of the LO and NLO weyl transformations, such that $\omega \rightarrow l \omega$ and $\sigma \rightarrow  l\sigma$. From \eqref{eq:Bdy_trans1}-\eqref{eq:Bdy_trans2} it is now easy to see we get a conformal class of type II Newton-Cartan geometries on the boundary.

Having worked out the transformations of the boundary vielbein, we are now able to compute boundary symmetries as well. So let us do that for a very simple example, namely global non-relativistic AdS. In this case the tetrad is given by
\begin{align}
    \tau = \frac{l}{r} dt\,, \qquad e^a =\frac{l}{r} dx^a\,,  \qquad e^{d+1} = \frac{l}{r} dr\,,
\end{align}
where $m =0$ and $\pi^a =0$. It follows that
\begin{align}
    \tau_\partial = \bar \tau = dt\,, \qquad e^a_\partial = \bar e^a = dx^a\,, \qquad \bar e^{d+1} = dr\,.
\end{align}
If we then require that 
\begin{align} \label{eq:BdyKillingSym}
    \delta \tau_\partial =0\,, \qquad  \delta e^a_\partial =0\,, \qquad \delta m_\partial =0\,, \qquad \delta \pi^a_\partial =0\,,
\end{align}
we find, as expected, the same infinite lift of the finite dimensional GCA as what we got in (\ref{eq:GCAlift}) but realised as conformal Killing symmetries instead.

\section{Excitation with planar symmetries} \label{sec:Excitation&planarSymmetries}
In this section we construct the non-relativistic AdS brane solution to NRG. We do this first by constructing an
ansatz that is asymptotically NR AdS and has the appropriate planar symmetries. We then substitute the ansatz into the NRG equation of motion and find that the solution corresponds to the $1/c^2$ expansion of the relativistic AdS black brane. Next, in section \ref{sec:GalBoost} we  perform a Galilean boost of this solution that leaves the boundary NC geometry invariant in order to introduce a velocity to our boundary EMT. We then compute the non-relativistic boundary EMT by using the $1/c^2$-expanded expression for the relativistic EMT. This way we bypass holographic renormalisation by assuming that it commutes with the $1/c^2$-expansion. Finally, in section \ref{sec:DualFluid} we see that due to the conformal symmetry of the boundary we get a massless Galilean fluid which is characterised by a vanishing momentum density. 

\subsection{Static solution} \label{sec:4.1}
In order to get a fluid on the boundary we want an asymptotically non-relativistic AdS solution with planar symmetries. This means that we want the LO geometry to be given by \eqref{eq:NRAdS2} and then we want to write down the most general $\Phi_{\mu\nu}$ (from the way $\Phi_{\mu\nu}$ appears in the NRG action we see that we only need its spatial components) and $m_\mu$ that respect time translation, boundary space translation and boundary rotation symmetries up to a gauge transformation. To this end it is useful to construct a gauge invariant curvature for these fields, implement the symmetry requirements on the curvature and then to solve these curvature conditions for the gauge fields by fixing the gauge.

The details of this analysis can be found in appendix \ref{app:ansatz}\footnote{In appendix \ref{app:ansatz}, the calculation is carried out for $d=3$ but we very strongly suspect that this result holds for general dimensions. Therefore, we apply the result in this section where we keep $D$ arbitrary.}. The final result is that imposing time translation, boundary space translation and boundary rotation invariance leads to
\begin{eqnarray}
    &&\Phi_{ij}=0\,,\qquad\Phi_{ri}=0\,,\qquad\Phi_{rr}=\Phi_{rr}(r)\,,\\
    && m_i=0\,,\qquad m_r=0\,,\qquad m_t=m_t(r)\,,
\end{eqnarray}
up to a gauge transformation.

We are then ready to solve the NRG EOM for this ansatz. Equations \eqref{eq:diva} and \eqref{eq:NLO-EOMhh} are solved by the LO geometry where
\begin{equation}
    \Lambda=-\frac{(D-1)(D-2)}{2l^2}\,.
\end{equation}
It follows from the ansatz that $h^{\rho\mu}h^{\sigma\nu}F_{\mu\nu}=0$, so using this along with \eqref{eq:vacRiem} we see that \eqref{eq:PvNLOEOM} is identically satisfied. Meanwhile, the Poisson equation in \eqref{eq:Pois} simplifies to
\begin{equation}
    e^{-1}\partial_\mu\left(e\tilde X^\mu\right)
    = -\frac{2}{D-2}\Lambda I\,, \label{eq:Poisansatz}
\end{equation}
where using the ansatz, we see that $\tilde X^\mu$ reduces to
\begin{eqnarray}
    \tilde X^\mu & = &  -h^{\mu\sigma}v^\rho \partial_\sigma m_\rho -\frac{1}{2}h^{\mu\kappa}h^{\nu\lambda}\left(a_\nu\Phi_{\lambda\kappa}+a_\lambda\Phi_{\nu\kappa}-a_\kappa\Phi_{\nu\lambda}\right)\,.
\end{eqnarray}
Thus, equation \eqref{eq:Poisansatz} reduces to
\begin{equation}\label{eq:1}
    r^2\partial_r^2 m_t-(D-3)r\partial_r m_t-(D-1)m_t+\frac{r}{2l}\left(r\partial_r\Phi_{rr}-2(D-2)\Phi_{rr}\right)=0\,.
\end{equation}

Finally, equation \eqref{eq:NNLO-EOMhh} becomes
\begin{equation}\label{eq:varhsimple}
    h^{\mu\alpha}h^{\nu\beta}\left(\check\nabla_\lambda\bar\chi^\lambda_{\mu\nu}-\frac{1}{2}\check\nabla_\mu\check\nabla_\nu \left(h^{\kappa\lambda}\Phi_{\kappa\lambda}\right)+\bar\chi^\rho_{\mu\nu}a_\rho+v^\rho\left(\check\nabla_{(\mu}+2a_{(\mu}\right)F_{\nu)\rho}+(D-1)l^{-2}\Phi_{\mu\nu}\right)=0\,.
\end{equation}
The nonzero components of $\bar\chi^\rho_{\mu\nu}$ are
\begin{eqnarray}
    \bar\chi^r_{ij} & = & -\delta_{ij}\frac{r}{l^2}\Phi_{rr}\,,\\
    \bar\chi^r_{rr} & = & \frac{r^2}{2l^2}\partial_r\Phi_{rr}+\frac{r}{l^2}\Phi_{rr}\,.
\end{eqnarray}
The nonzero components of \eqref{eq:varhsimple} are the $ij$ and $rr$ components where the $ij$ components are pure trace. More specifically, for the $ij$ component we get
\begin{equation}\label{eq:2}
    \frac{r}{2l}\left(r\partial_r\Phi_{rr}-2(D-2)\Phi_{rr}\right)-r\partial_r m_t-m_t=0\,.
\end{equation}
Combining this with \eqref{eq:1} leads to
\begin{equation} \label{eq:2.5}
    r^2\partial_r^2 m_t-(D-4)r\partial_r m_t-(D-2)m_t=0\,.
\end{equation}
Finally, the $rr$ component gives
\begin{equation}\label{eq:3}
    r^2\partial_r^2 m_t+r\partial_r m_t-m_t+\frac{1}{2l}(D-1)r^2\partial_r\Phi_{rr}=0\,.
\end{equation}
Solving equations \eqref{eq:2.5}--\eqref{eq:3} gives
\begin{eqnarray}
    m_t & = & l^{-1}Ar^{D-2}+Br^{-1}\,,\\
    \Phi_{rr} & = & -2A r^{D-3}\,,
\end{eqnarray}
where $A$ and $B$ are constants. We can use the residual gauge transformation of our gauge choice $m_i=m_r=0$, i.e. \eqref{eq:resLambda}, to set $B=0$.

In order to write the solution in the same gauge as we used for the definition of the boundary we notice that the transformation under NLO diffeomorphisms can be trivially exponentiated to a finite transformation. So under $\zeta^\mu$ we have
\begin{eqnarray}
    m'_\mu & = & m_\mu+\mathcal{L}_\zeta\tau_\mu\,,\\
    \Phi'_{\mu\nu} & = & \Phi_{\mu\nu}+\mathcal{L}_\zeta h_{\mu\nu}\,.
\end{eqnarray}
The reason for this is that $\tau_\mu$ and the parameter $\zeta^\mu$ do not transform under subleading diffeomorphisms so when we exponentiate the transformation it stops at the linear order. We can use this to set $\Phi'_{rr}=0$ by choosing 
\begin{equation}
    \zeta^r=\frac{1}{l^2}\frac{A}{D-1}r^D\,,\qquad \zeta^i=\zeta^t=0\,.
\end{equation}
In this new gauge we have the nonzero components
\begin{eqnarray}
    m'_t & = & \frac{D-2}{D-1}\frac{1}{l}Ar^{D-2}\,,\\
    \Phi'_{ij} & = & -\frac{2}{D-1}Ar^{D-3}\delta_{ij}\,.
\end{eqnarray}

If we compare with the $1/c^2$ expansion of the AdS black brane (see equation \eqref{eq:mt}) we see that we have
\begin{equation}
    A=-\frac{8\pi G}{D-2}l^{4-D}m\,.
\end{equation}

\subsection{Galilean boosted solution} \label{sec:GalBoost}

We will now boost our solution to introduce the fluid velocity. The boost should be an asymptotic Killing symmetry, i.e. one that leaves the NC boundary geometry invariant. We also want this boost to be a finite transformation in order to get a finite velocity parameter.

More specifically, it is the fields $\tau_A$, $h_{AB}$ and $m_A$\footnote{The reason we do not include the boundary $\Phi_{AB}$ to this list is that the response to this field is the NLO momentum current which we do not care about at this order. Nonetheless, one could have still required that the boundary $\Phi_{AB}$ was invariant and it would not have changed the final result for the LO boundary EMT. It would just be a longer calculation where one would have to include NLO local boosts as well.} evaluated on the boundary (whose responses are the energy, momentum and mass currents) that we want to stay invariant under this transformation. Thus, we want to perform a local Galilean boost ($\lambda_\mu$), a NLO diffeomorphism ($\zeta^\mu$) and a LO diffeomorphism that fulfills the first three conditions in \eqref{eq:BdyKillingSym} and takes the form of a finite Galilean boost. This finite transformation can most easily be described in two steps. First a finite NLO diffeomorphism and local Galilean boost with parameters
\begin{eqnarray}
    \lambda_\mu & = & \frac{l}{r}v^i\delta_\mu^i\,,\\
    \zeta^\mu & = & -\left(v^i x^i+\frac{1}{2}v^2 t\right)\delta^\mu_t\,,
\end{eqnarray}
where $v^i$ is a constant, which act as\footnote{In order to compute the finite transformation of the boost we made use of \eqref{eq:traofh}-\eqref{eq:traofm}, as well as the fact that $\lambda_\mu = \lambda_a e^a_\mu$ where $\delta_\lambda e^a_\mu =\lambda^a \tau_\mu $.}
\begin{eqnarray}
    h'_{\mu\nu} & = & h_{\mu\nu}+\tau_\mu\lambda_\nu+\tau_\nu\lambda_\mu+\lambda^2\tau_\mu\tau_\nu\,,\\
    m'_\mu & = & m_\mu+\mathcal{L}_\zeta\tau_\mu+\lambda_\mu+\frac{1}{2}\lambda^2\tau_\mu\,.
\end{eqnarray}
The second step is then to perform the following LO diffeomorphism
\begin{equation}
    t' =t\,,\qquad x'^i=x^i+v^it\,.
\end{equation}
This leads to 
\begin{eqnarray}
    \tau' & = & \frac{l}{r}dt'\,,\\
    h' & = & \frac{l^2}{r^2}\left(dr^2+d\vec x'^2\right)\,,\\
    m' & = & \frac{D-2}{D-1}\frac{1}{l}Ar^{D-2} dt'\,.
\end{eqnarray}
So the transformations have actually left the whole of $\tau, h, m$ invariant. 

However, more subleading fields in the $1/c^2$ expansion will get affected. In particular the field $\bar\Phi_{\mu\nu}$ (see definition in \eqref{eq:Phibar}), which is inert under the local Galilean boosts, transforms into (under the NLO and LO diffeomorphisms) 
\begin{eqnarray}
    \bar\Phi' & = & \bar\Phi_{\mu\nu}dx^\mu dx^\nu+\mathcal{L}_\zeta h_{\mu\nu}dx^\mu dx^\nu\nonumber\\
    & = & \left(\bar\Phi_{tt}+2\bar h_{tt}\partial_t\zeta^t\right)dt^2+2\left(\bar\Phi_{ti}+\bar h_{tt}\partial_i\zeta^t\right)dt dx^i+\bar\Phi_{ij}dx^i dx^j\nonumber\\
    & = & \bar\Phi'_{t't'}dt'^2+2\bar\Phi'_{t'i}dt' dx'^i+\bar\Phi'_{ij}dx'^i dx'^j\,,
\end{eqnarray}
where
\begin{eqnarray}
    \bar\Phi'_{t't'} & = & \bar\Phi_{tt}+2\bar h_{tt}\partial_t\zeta^t-2v^i\left(\bar\Phi_{ti}+\bar h_{tt}\partial_i\zeta^t\right)+v^i v^j\bar\Phi_{ij}\,,\\
    \bar\Phi'_{t'i} & = & \bar\Phi_{ti}+\bar h_{tt}\partial_i\zeta^t-v^j\bar\Phi_{ij}\,,\\
    \bar\Phi'_{ij} & = & \bar\Phi_{ij}\,.
\end{eqnarray}
In FG coordinates we thus get
\begin{eqnarray}
        \bar{h}'_{\mu \nu} dx'^\mu dx'^\nu &=& \frac{l^2}{r^2} \big( dr^2 + d \vec{x}'^2 \big) + \frac{16\pi G}{D-1}l^{4-D}m r^{D-3} dt'^2 \,,
        \\
        \bar{\Phi}'_{\mu \nu} dx'^\mu dx'^\nu& =& \frac{16 \pi G}{D-1}l^{4-D}m r^{D-3} \left( \frac{1}{D-2} d\vec{x}'^2  - 2\frac{D-1}{D-2} v_i dx'^i dt'+\frac{D-1}{D-2} v^2 dt'^2\right) \nonumber\\
        &&+\bar\Phi_{tt}dt'^2+2\bar\Phi_{ti}dt' dx'^i \label{eq:barPhi}\,.
\end{eqnarray}

For zero velocity there is only one scale in the game that controls the deformation, i.e. $m$. There is one dimensionless ratio namely $l/L$ with $L^{-1}\sim mG/c^2$ which is order $c^{-2}$. The near boundary $r$ expansion is an expansion in $r/l$ where $l$ is the curvature radius. Consider $T_\mu$ and $\Pi_{\mu\nu}$. N$^n$LO means $\mathcal{O}((l/L)^n)$ so the NNLO fields that contribute to $\bar{\Phi}_{tt}$ and $\bar{\Phi}_{ti}$ are $\mathcal{O}((l/L)^2)$. There is a scaling argument\footnote{When $m=0$ there is a dilatation symmetry that rescales $t, x^i, r$ all in the same way. When $m\neq 0$ this symmetry is broken but in such a way that a dilatation is equivalent to rescaling only $m$ to an appropriate dimension dependent power.} that tells us that we  have an expansion in 
\begin{equation*}
    \frac{l}{L}\left(\frac{r}{l}\right)^{D-1}\,.
\end{equation*}
This tells us that the fields on the second line in \eqref{eq:barPhi} are order $m^2$ and thus more subleading in $r$ than the terms on the first line. For this reason the terms on the second line in \eqref{eq:barPhi} will prove irrelevant in the computation of the leading order boundary EMT.

Usually, in order to identify the boundary EMT one has to go through the process of holographic renormalisation. However, since we expect this process to commute with the $1/c^2$-expansion, we have taken the relativistic expression for the boundary EMT and $1/c^2$-expanded it to find an expression for the non-relativistic boundary EMT 
 in terms of the NR fields (see appendix \ref{App:NRExpansionOfAsympAdS}). More specifically, using equations \eqref{eq:EMTinNRMet1}-\eqref{eq:EMTinNRMet2} along with what we found above we get (for $D=5$)
 \begin{align}
        \mathcal{T}^t_{\ t} =  -m\,,
        \qquad
         \mathcal{T}^i_{\ t}= - \frac{4}{3}m v^i\,,
        \qquad
        \mathcal{T}^t_{\ i} =0\,,\qquad
        \mathcal{T}^i_{\ j} = \frac{m}{3} \delta^i_j\,.
    \end{align}
Additionally, we expect that the result in \eqref{eq:EMTinNRMet1}-\eqref{eq:EMTinNRMet2} to be straightforwardly generalisable to arbitrary dimensions in which case we would get
\begin{align}\label{eq:BoundryEMT}
            \mathcal{T}^t_{\ t} =  -m\,,
        \qquad
         \mathcal{T}^i_{\ t}= - \frac{D-1}{D-2}m v^i\,,
        \qquad
        \mathcal{T}^t_{\ i} =0\,,\qquad
        \mathcal{T}^i_{\ j} = \frac{m}{D-2} \delta^i_j\,.
\end{align}

Lastly, we should note that components such as $\bar\Phi_{t' i}$ are components that did not appear in the NRG theory. However, we were able to compute the relevant parts of these fields from the NRG fields and basic transformation rules.
Nonetheless, the fact that (components of) a field that is not in the theory shows up in the expression for the EMT seems problematic. The underlying reason for this apparent inconsistency is that we avoided holographic renormalisation and instead relied on the expansion of the relativistic expression of the EMT. The expanded result would have been consistent with the non-relativistic theory had we included the full NNLO action but since we used the NRG action the field components $\bar\Phi_{t' i}$ have been swapped out for $\zeta_{\mu \nu}$, which leads to this situation. This can be resolved either by considering the full NNLO action in which case $\bar\Phi_{t' i}$ would be part of the theory or by doing holographic renormalisation of the NRG action in which case it would be components of $\zeta_{\mu \nu}$, instead of $\bar\Phi_{t' i}$, that would contain components of the boundary EMT. In either case it does not change the end result for the LO boundary EMT in equation \eqref{eq:BoundryEMT}.

\subsection{Dual Galilean fluid} \label{sec:DualFluid}

Next, we show that this result makes sense from the point of the view of the $1/c^2$ expansion of a conformal relativistic fluid.

Let us first make some general comments about the $1/c^2$ expansion of relativistic perfect fluids in order to understand the type of fluid we will get on the boundary. There are essentially two types of Galilean fluids, Bargmann (massive Galilean) fluids and massless Galilean fluids. The former is well-known and describes many of the fluids we see in everyday life. The latter fluid is much less understood but still potentially very interesting in its own right (see e.g. \cite{deBoer:2017ing,Ciambelli:2018xat,Hansen:2020pqs,Petkou:2022bmz}). For more on massless Galilean particles\footnote{There are two kinds of massless Galilean particles called electric and magnetic in \cite{Bergshoeff:2022qkx}.} see e.g. \cite{Hansen:2019vqf,Hansen:2020pqs,Bergshoeff:2022qkx}. We will see that the equation of state due to conformal symmetry forces us to consider massless Galilean fluids.

The stress-energy tensor of a relativistic perfect fluid is given by
\begin{align}
    T^\mu_{\ \nu} = \frac{E+P}{c^2} U^\mu U_\nu + P \delta^\mu_\nu\,,
\end{align}
where $P$ is the pressure, $E$ is the energy and $U^\mu = \gamma (1,\vec{v})$ the velocity. Furthermore, $U_\nu=\eta_{\nu\rho}U^\rho=\gamma(-c^2,\vec v)$. The equation of state of a relativistic conformal fluid in $D-1=d+1$ dimensions is $E=(D-2)P$. Hence, the $1/c^2$ expansion of $E$ and $P$ have to start at the same order in $c^2$. This is not what happens in the Bargmann limit in which case $E=mc^2+E_0+\mathcal{O}(c^{-2})$ and $P=P_0+\mathcal{O}(c^{-2})$. See for example \cite{Hansen:2020pqs}. 

In components, the relativistic EMT is 
\begin{eqnarray}
    T^t{}_t & = & -\gamma^2 E+(1-\gamma^2)P\,,\\
    T^t{}_j & = & c^{-2}\gamma^2(E+P)v^j\,,\\
    T^i{}_t & = & -\gamma^2(E+P)v^i\,,\\
    T^i{}_j & = & P\delta^i_j+c^{-2}(E+P)v^i v^j\,.
\end{eqnarray}
We have $\gamma=1+\mathcal{O}(c^{-2})$. We thus get the following expression for the LO EMT
\begin{eqnarray}
    T^t{}_t & = & -\mathcal{E}\,,\\
    T^t{}_j & = & 0\,,\\
    T^i{}_t & = & -\frac{D-1}{D-2}\mathcal{E}v^i\,,\\
    T^i{}_j & = & \frac{1}{D-2}\mathcal{E}\delta^i_j\,,
\end{eqnarray}
where we used the equation of state to eliminate $P$, and where we used $\mathcal{E}$ to denote the LO term in the expansion of $E$, which may include a factor of $c$ if the LO term is not order $c^0$ (for example for the AdS black brane $\mathcal{E}=mc^2$). This is of the same form as \eqref{eq:BoundryEMT}.

Since the momentum is zero we call this a massless Galilean fluid \cite{deBoer:2017ing}. For a Bargmann fluid the momentum is equal to the mass flux, so in the limit where the mass goes to zero we obtain a fluid with vanishing momentum. This follows from Galilean boost invariance. We thus conclude that the boundary description of our boosted solution is a massless Galilean fluid.

\section{Discussion}
The next step in the endeavour of constructing  a non-relativistic fluid/gravity correspondence is to perturb the boundary fluid away from global equilibrium by promoting the velocity and mass parameter of the non-relativistic AdS brane to local functions of $(t,x^i)$. However, this comes with a subtlety associated with ensuring regularity in the bulk for the perturbed spacetime. In the relativistic case, Schwarzschild coordinates are ill-equipped to deal with these issues \cite{Rangamani:2009xk} and so the standard approach is to work in Eddington-Finkelstein coordinates. It is not immediately clear how these problems translate to the non-relativistic AdS brane because there is no obvious notion of a horizon. Essentially, one wants to ensure that the bulk solution is regular when the boundary fluid equations are obeyed. 

Aside from the most natural next step, this project has also led to several other interesting questions and follow-ups. Firstly, it would be interesting to better understand the dual massless Galilean fluid on the boundary. This includes questions of whether a massless Galilean fluid can be derived from some microscopic description and which types of physical systems it can effectively describe. It would also be very interesting to see if we can try to change the setup in such a way that we end up with a massive Galilean (Bargmann) fluid on the boundary.

In terms of the bulk description, it is interesting to wonder how many properties the non-relativistic AdS brane solution inherits from the relativistic parent geometry. First, it should be noted that it is quite deliberately not called a non-relativistic AdS {\it black} brane because there is no notion of a horizon. Nothing special happens at $R=1/b$ and the solution is perfectly regular at every point for which $R>0$. Whether it is a good approximation of the relativistic solution is a different question and here it inevitably fails to be so at $R=1/b$. However, one can still wonder whether the non-relativistic AdS brane can be thought of as a proper thermodynamic object in the same way as its parents geometry. More specifically, whether one can ascribe temperature and entropy to it. For this one could draw inspiration from what was done for so-called Carrollian black holes \cite{Ecker:2023uwm}, but it should be stressed that the situation here is different from \cite{Ecker:2023uwm} in that the LO geometry here is everywhere regular.

In this paper we have also presented new results for the theory of non-relativistic gravity more broadly. One of these results was the simplification of the EOM for the NRG action. This came out of a rewriting inspired by the comparison with the expansion of Einstein's field equations. Now, one can wonder if there is a rewriting of the action itself for which the variation gives the simplified EOM manifestly or perhaps there is a better way to perform the variations that immediately gives the simplified EOM. It would also be interesting to write the equations \eqref{eq:diva}--\eqref{eq:NNLO-EOMhh} for general matter couplings.

Lastly, one could consider the question of whether there exists a fluid/gravity correspondence in the opposite limit, $c \rightarrow 0$. More precisely, one could go through the same steps of 
expanding GR with a negative cosmological constant around $c=0$ \cite{Hansen:2021fxi} which should lead to Carroll gravity  \cite{Hansen:2021fxi,Figueroa-OFarrill:2022mcy,Hartong:2015xda} in the bulk, and then trying to construct similar solutions as the ones found in this paper but using Carroll geometry.

\addcontentsline{toc}{section}{Acknowledgements}
\section*{Acknowledgements}

JH was supported by the Royal Society University Research Fellowship
Renewal “Non-Lorentzian String Theory” (grant number URF\textbackslash
R\textbackslash 221038). AM was in part supported by the Royal Society through the Enhanced Research Expenses 2021 Award (grant number RF\textbackslash ERE\textbackslash 210139).

\appendix

\section{Affine connection, torsion and curvature}\label{app:curvature}

In this appendix we have collected some helpful formulas regarding affine connections with nonzero torsion, like $\check{\Gamma}^\rho_{\mu \nu} =- v^\rho\partial_\mu\tau_\nu+\frac{1}{2}h^{\rho\sigma}\left(\partial_\mu h_{\nu\sigma}+\partial_\nu h_{\mu\sigma}-\partial_\sigma h_{\mu\nu}\right)$ which is used throughout the paper. We also cover the special properties of the $\check{\Gamma}^\rho_{\mu \nu}$-connection. 

We denote covariant derivatives by $\nabla_\mu$, the Riemann tensor by ${R}_{\mu\nu\sigma}{}^{\rho}$, and the torsion tensor by $T^\rho{}_{\mu\nu}$. The Riemann and the torsion tensors are defined by
\begin{eqnarray}
\left[\nabla_\mu,\nabla_\nu\right]X_\sigma & = & R_{\mu\nu\sigma}{}^\rho X_\rho-T^\rho{}_{\mu\nu}\nabla_\rho X_\sigma\,,\\
\left[\nabla_\mu,\nabla_\nu\right]X^\rho & = & -R_{\mu\nu\sigma}{}^\rho X^\sigma-T^\sigma{}_{\mu\nu}\nabla_\sigma X^\rho\,.
\end{eqnarray}
It follows from these equations that
\begin{eqnarray}
{R}_{\mu\nu\sigma}{}^{\rho}&\equiv&\partial_{\nu}{\Gamma}_{\mu\sigma}^{\rho}-\partial_{\mu}{\Gamma}_{\nu\sigma}^{\rho}-{\Gamma}_{\mu\lambda}^{\rho}{\Gamma}_{\nu\sigma}^{\lambda}+{\Gamma}_{\nu\lambda}^{\rho}{\Gamma}_{\mu\sigma}^{\lambda}\label{eq:Riemann_tensor}\,,\\
T^\rho{}_{\mu\nu} &\equiv& 2\Gamma^\rho_{[\mu\nu]}\,.\label{eq:torsion_tensor}
\end{eqnarray}
The algebraic and differential Bianchi identities are then given by 
\begin{eqnarray}
    R_{[\mu\nu\sigma]}{}^\rho & = & T^\lambda{}_{[\mu\nu}T^\rho{}_{\sigma]\lambda}-\nabla_{[\mu}T^\rho{}_{\nu\sigma]}\,,\\
    \nabla_{[\lambda}R_{\mu\nu]\sigma}{}^\kappa & = & T^\rho{}_{[\lambda\mu}R_{\nu]\rho\sigma}{}^\kappa\,.\label{eq:diffBianchi}
\end{eqnarray}
The Ricci tensor is defined as
\begin{equation}\label{eq:Ricci_tensor_LC}
R_{\mu\nu} \equiv {R}_{\mu\rho\nu}{}^{\rho}\,.
\end{equation}

The connection we choose to work with in paper is $\Check{\Gamma}^\rho_{\mu \nu}$. This connections satisfies the property that
\begin{equation}
    \check \Gamma^{\rho}_{\mu\rho}=\partial_\mu\log e\,,
\end{equation}
where $e=\text{det}(\tau_\mu\,, e^a_\mu)$ is the integration measure. It then follows that
\begin{equation}
R_{\mu\nu\rho}{}^\rho=0\,.
\end{equation}
We then have an non-zero antisymmetric part of the Ricci tensor. This is given by
\begin{equation}\label{eq:ASpartRicci}
    2R_{[\mu\nu]}=T^\lambda{}_{\mu\nu}T^\rho{}_{\lambda\rho}+
    \nabla_\mu T^\rho{}_{\nu\rho}-\nabla_\nu T^\rho{}_{\mu\rho}+\nabla_\rho T^\rho{}_{\mu\nu}\,,
\end{equation}
where for $\check{\Gamma}^\rho_{\mu \nu}$ the torsion is given by $T^\rho{}_{\mu \nu} = -v^\rho \tau_{\mu \nu}$ with $\tau_{\mu \nu}=\partial_\mu\tau_\nu-\partial_\nu\tau_\mu$. 

Finally, we will derive the most general Newton-Cartan metric compatible connection that is made entirely of LO fields ($\tau_\mu$ and $h_{\mu\nu}$) and derivatives thereof, and whose torsion is equal to the intrinsic torsion \cite{Figueroa-OFarrill:2020gpr}, i.e. proportional to $\tau_{\mu\nu}$. By Newton-Cartan metric compatibility we specifically mean
\begin{align}
    \nabla_\mu \tau_\nu = 0\,,
    \qquad
    \nabla_\mu h^{\nu \rho} =0\,.
\end{align}

Firstly, the torsion requirement means that we can write 
\begin{align}
        \Gamma^\rho_{\mu \nu} &= \check{\Gamma}^\rho_{\mu \nu} + C^\rho_{\mu \nu}\,,
\end{align}
where $C^\rho_{[\mu \nu]}=0$. From here on one can follow the same argument as in section 3.2 of \cite{Hartong:2022lsy} to see that the metric compatibility condition leads to 
\begin{align}
    C^\rho_{\mu \nu} = - \frac{1}{2} h^{\rho \sigma} \left( \tau_\mu F_{\nu \sigma} + \tau_\nu F_{\mu \sigma} \right)\,,
\end{align}
where $F_{\mu \nu}$ is an arbitrary antisymmetric (0,2) tensor. The only way we can make an antisymmetric (0,2) tensor out of the LO fields is if $F_{\mu \nu} \propto \tau_{\mu \nu}$. We then use that for TTNC spacetimes $\tau_{\mu \nu} = 2a_{[\mu} \tau_{\nu]}$, such that 
\begin{align}
    C^\rho_{\mu \nu} \propto h^{\rho \sigma} a_\sigma \tau_\mu \tau_\nu\,.
\end{align}
So, in the end we find that the most general Newton-Cartan metric compatible connection that is made entirely of LO fields ($\tau_\mu$ and $h_{\mu\nu}$) and whose the torsion is equal to the intrinsic torsion of a TTNC spacetime is given by
\begin{align}
        \Gamma^\rho_{\mu \nu} &= \check{\Gamma}^\rho_{\mu \nu} + \alpha h^{\rho \sigma} a_\sigma \tau_\mu \tau_\nu\,,
\end{align}
where $\alpha$ is an arbitrary real number. 

\section{Equations of motion of non-relativistic gravity}\label{app:EOM}
This appendix serves two main purposes. Firstly, we show that the EOM of the NRG action are equivalent to the expansion of Einstein's field equations, this is shown in \ref{app:ComparingWithOn-shell}. The second purpose is to correct and simplify the NRG equations of motion given in \cite{Hansen:2020pqs}. The correction involves a missing factor of $-1/2$ in one term and a wrong placement of indices in another. The corrected equation can be found in \eqref{eq:NRG_EOM4}. The simplification of the NRG EOM involves recombining the original equations of motion, \eqref{eq:NRG_EOM1}-\eqref{eq:NRG_EOM4}, in a manner that significantly reduces the size of especially the last equation. The final form of the EOM are given in section \ref{app:Simplification}. This final rewriting of the NRG EOM was inspired by the expansion of Einstein's field equations which come out in a more compact form than \eqref{eq:NRG_EOM1}-\eqref{eq:NRG_EOM4}.

\subsection{Variations and Ward identities} \label{app:WardIden}
We define the response to varying the NRG Lagrangian in \eqref{eq:action_NRG} with respect to the LO and NLO fields as
\begin{equation}\label{eq:EOMs_NRG_variation}
\delta\mathcal{L_{\mathrm{NRG}}} \equiv -\frac{e}{8\pi G_N} \left(\mathcal{G}_{\tau}^\mu \delta \tau_\mu + \mathcal{G}_{m}^\mu \delta m_\mu
+\frac{1}{2}\mathcal{G}_{h}^{\mu\nu}\delta h_{\mu\nu}+\frac{1}{2}\mathcal{G}_{\Phi}^{\mu\nu}\delta{\Phi}_{\mu\nu}\right)\,.
\end{equation}
We next derive some Ward identities (WI) that follow from the local symmetries of NRG. Under a local Galilean boost with parameter $\lambda_\mu$ (obeying $v^\mu\lambda_\mu=0$) the NRG fields transform as
\begin{equation}
\delta h_{\mu\nu} = \lambda_\mu\tau_\nu+\lambda_\nu\tau_\mu\,,\qquad \delta m_\mu = \lambda_\mu\,,\qquad
\delta\Phi_{\mu\nu} = \lambda_\mu m_\nu+\lambda_\nu m_\mu\,.
\end{equation}
The field $\Phi_{\mu\nu}$ also transforms under a NLO Galilean boost with parameter $\kappa_\mu$ (see eq. \eqref{eq:trafoPhi}) but this leaves the NRG Lagrangian invariant because we have $\tau_\mu\mathcal{G}_{\Phi}^{\mu\nu}=0$. The WI for Galilean boost invariance is thus
\begin{equation}\label{eq:WIGalboosts}
    P_\mu^\rho\left(\mathcal{G}_m^\mu+\tau_\nu\mathcal{G}_h^{\mu\nu}+m_\nu\mathcal{G}_\Phi^{\mu\nu}\right)=0\,.
\end{equation}
We can use this to our benefit and read this equation as providing $P_\mu^\rho\tau_\nu\mathcal{G}_h^{\mu\nu}$. The benefit of doing this is that when we vary $h_{\mu\nu}$ we only have to consider its spatial part, i.e.
$\delta_P h_{\rho\sigma}=P_\rho^\mu P_\sigma^\nu\delta h_{\mu\nu}\,$.
To see this we note that 
\begin{equation}
    \frac{1}{2}\mathcal{G}_h^{\mu\nu}\delta h_{\mu\nu}=\frac{1}{2}\mathcal{G}_h^{\rho\sigma}P_\rho^\mu P_\sigma^\nu\delta h_{\mu\nu}- \mathcal{G}_h^{\rho\sigma}\tau_\rho P_\sigma^\nu v^\mu\delta h_{\mu\nu}\,,
\end{equation}
where we note that $v^\mu v^\nu\delta h_{\mu\nu}=0$. The second term in this variation is given by the boost Ward identity and so we can restrict our attention to $\delta_P h_{\rho\sigma}$.
Furthermore, as argued above in order to avoid equations containing the Lagrange multiplier $\zeta_{\mu\nu}$ we only vary $\tau_\mu$ as $\delta\tau_\mu=\Omega\tau_\mu$. 

The NRG equations of motion are given in \cite{Hansen:2020pqs} and we will restate (a corrected version of) them here
\begin{eqnarray}
\tau_\mu \mathcal{G}_{\tau}^\mu &=& -\frac{1}{2}\Big[\left(h^{\mu\nu}K_{\mu\nu}\right)^2-h^{\mu\rho}h^{\nu\sigma}K_{\mu\nu}K_{\rho\sigma}+\frac{3}{4}h^{\mu\rho}h^{\nu\sigma}F_{\mu\nu}F_{\rho\sigma}\nonumber \\
&&+m_\nu\left[\left(\check\nabla_\mu+2a_\mu\right)h^{\mu\rho}h^{\nu\sigma}F_{\rho\sigma}-2\left(h^{\mu\rho}h^{\nu\sigma}-h^{\mu\nu}h^{\rho\sigma}\right)\check\nabla_\mu K_{\rho\sigma}\right]\nonumber\\
&&-2\left(h^{\mu\rho}h^{\nu\sigma}-h^{\mu\nu}h^{\rho\sigma}\right)K_{\rho\sigma}\left(\check\nabla_\mu+a_\mu\right)m_\nu
+\left(h^{\mu\rho}h^{\nu\sigma}-h^{\mu\nu}h^{\rho\sigma}\right)\check{\nabla}_\mu \check{\nabla}_\nu\Phi_{\rho\sigma}\nonumber \\
&&-\Phi_{\rho\sigma}h^{\mu\rho}h^{\nu\sigma}\left(\check R_{\mu\nu}-\frac{1}{2}h_{\mu\nu}h^{\kappa\lambda}\check R_{\kappa\lambda}\right)\Big]\,,\label{eq:NRG_EOM1}\\
\mathcal{G}_{m}^\nu &=& \frac{1}{2}\Big[ 2\left(h^{\mu\rho}h^{\nu\sigma}-h^{\mu\nu}h^{\rho\sigma}\right)\check{\nabla}_\mu K_{\rho\sigma}
+v^\nu h^{\mu\rho}\check R_{\mu\rho}
+\left(\check{\nabla}_\mu+2a_\mu \right)h^{\mu\rho}h^{\nu\sigma}F_{\rho\sigma}\Big]\,,\label{eq:NRG_EOM2}\\
\mathcal{G}_{\Phi}^{\rho\sigma}&=&h^{\mu\rho}h^{\nu\sigma}\left(\check R_{\mu\nu}
-\frac{1}{2}h_{\mu\nu}h^{\kappa\lambda}\check R_{\kappa\lambda}-\left(\check{\nabla}_\mu +a_\mu\right)a_\nu 
+h_{\mu\nu}h^{\kappa\lambda} \left(\check\nabla_{\kappa}+a_\kappa\right)a_\lambda\right)\,.\label{eq:NRG_EOM3}
\\
\mathcal{G}_{h}^{\mu\nu}P^\alpha_\mu P^\beta_\nu &=& -\frac{1}{2}h^{\alpha\beta}\left(h^{\mu\rho}h^{\nu\sigma}-h^{\mu\nu}h^{\rho\sigma}\right)K_{\mu\nu}K_{\rho\sigma}-\check\nabla_\lambda\left(v^\lambda\left(h^{\mu\alpha}h^{\nu\beta}-h^{\alpha\beta}h^{\mu\nu}\right)K_{\mu\nu}\right)\nonumber\\
&&-h^{\alpha\beta}h^{\mu\rho}h^{\nu\sigma}K_{\rho\sigma}\left(\check\nabla_\mu+a_\mu\right)m_\nu+h^{\alpha\beta}\check\nabla_\lambda\left(v^\lambda h^{\mu\nu}\left(\check\nabla_\mu+a_\mu\right)m_\nu\right)\nonumber\\
&&-\frac{1}{2}\check\nabla_\lambda\left(v^\lambda\left(h^{\mu\alpha}h^{\nu\beta}+h^{\nu\alpha}h^{\mu\beta}\right)\left(\check\nabla_\mu+a_\mu\right)m_\nu\right)\nonumber\\
&&+h^{\alpha\rho}h^{\beta\sigma}K_{\rho\sigma}h^{\mu\nu}\left(\check\nabla_\mu+a_\mu\right)m_\nu-\left(h^{\alpha\beta}h^{\rho\sigma}-h^{\alpha\rho}h^{\beta\sigma}\right)h^{\mu\nu}m_\mu\check\nabla_\nu K_{\rho\sigma}\nonumber\\
&&-h^{\mu\sigma}\left(h^{\alpha\rho}h^{\beta\nu}+h^{\beta\rho}h^{\alpha\nu}\right)m_\nu\left(\check\nabla_\mu K_{\rho\sigma}-\check\nabla_\rho K_{\mu\sigma}\right)\nonumber\\
&&-\frac{1}{2}\left(h^{\alpha\rho}h^{\beta\nu}+h^{\beta\rho}h^{\alpha\nu}\right)h^{\mu\sigma}K_{\rho\sigma}\left(F_{\mu\nu}+2a_\mu m_\nu-2a_\nu m_\mu\right)\nonumber\\
&&+\frac{1}{2}h^{\mu\alpha}h^{\rho\beta}h^{\nu\sigma}F_{\mu\nu}F_{\rho\sigma}-\frac{1}{8}h^{\alpha\beta}h^{\mu\rho}h^{\nu\sigma}F_{\mu\nu}F_{\rho\sigma}\nonumber\\
&&+\phi h^{\mu\alpha}h^{\nu\beta}\left(\check R_{\mu\nu}-\frac{1}{2}h_{\mu\nu}h^{\rho\sigma}\check R_{\rho\sigma}\right)+h^{\alpha\beta}h^{\rho\sigma}\left(\check\nabla_\rho+a_\rho\right)\left(\check\nabla_\sigma+a_\sigma\right)\phi\nonumber\\
&&-\frac{1}{2}\left(h^{\mu\alpha}h^{\rho\beta}+h^{\mu\beta}h^{\rho\alpha}\right)\left(\check\nabla_\mu+a_\mu\right)\left(\check\nabla_\rho+a_\rho\right)\phi\nonumber\\
&&-\frac{1}{2}\left(h^{\alpha\beta}h^{\mu\rho}h^{\nu\sigma}+h^{\mu\nu}h^{\alpha\rho}h^{\beta\sigma}-h^{\nu\sigma}\left(h^{\mu\alpha}h^{\rho\beta}+h^{\mu\beta}h^{\rho\alpha}\right)\right)\Phi_{\rho\sigma}\left(\check\nabla_\mu+a_\mu\right)a_\nu\nonumber\\
&&-\frac{1}{2}h^{\mu\rho}\left(h^{\nu\beta}h^{\sigma\alpha}+h^{\nu\alpha}h^{\sigma\beta}\right)a_\nu\check\nabla_\mu\Phi_{\rho\sigma}+\frac{1}{2}h^{\rho\sigma}\left(h^{\mu\alpha}h^{\nu\beta}+h^{\mu\beta}h^{\nu\alpha}\right)a_\nu\check\nabla_\mu\Phi_{\rho\sigma}\nonumber\\
&&+\frac{1}{2}h^{\mu\nu}\left(h^{\alpha\rho}h^{\beta\sigma}-h^{\alpha\beta}h^{\rho\sigma}\right)a_\nu\check\nabla_\mu\Phi_{\rho\sigma}+\frac{1}{2}h^{\kappa\lambda}\Phi_{\kappa\lambda}h^{\mu\alpha}h^{\nu\beta}\left(\check R_{\mu\nu}-\frac{1}{2}h_{\mu\nu}h^{\rho\sigma}\check R_{\rho\sigma}\right)\nonumber\\
&&-\frac{1}{4}h^{\kappa\lambda}\left(h^{\mu\alpha}h^{\rho\beta}+h^{\mu\beta}h^{\rho\alpha}\right)\left(\check\nabla_\mu+a_\mu\right)\left(\check\nabla_\rho+a_\rho\right)\Phi_{\kappa\lambda}\nonumber\\
&&+\frac{1}{2}h^{\rho\alpha}h^{\sigma\beta}\Phi_{\rho\sigma}h^{\mu\nu}\check R_{\mu\nu}+\frac{1}{2}h^{\kappa\lambda}h^{\alpha\beta}h^{\rho\sigma}\left(\check\nabla_\rho+a_\rho\right)\left(\check\nabla_\sigma+a_\sigma\right)\Phi_{\kappa\lambda}\nonumber\\
&&-\left(h^{\mu\alpha}h^{\rho\beta}h^{\nu\sigma}+h^{\mu\rho}h^{\nu\beta}h^{\sigma\alpha}\right)\Phi_{\rho\sigma}\check R_{\mu\nu}+\frac{1}{2}h^{\alpha\beta}h^{\mu\rho}h^{\nu\sigma}\Phi_{\rho\sigma}\check R_{\mu\nu}\nonumber\\
&&+\frac{1}{2}h^{\mu\rho}\left(h^{\alpha\sigma}h^{\beta\nu}+h^{\beta\sigma}h^{\alpha\nu}\right)\left(\check\nabla_\mu+a_\mu\right)\left(\check\nabla_\nu+a_\nu\right)\Phi_{\rho\sigma}\nonumber\\
&&-\frac{1}{2}\left(h^{\alpha\rho}h^{\beta\sigma}h^{\mu\nu}+h^{\nu\sigma}h^{\mu\rho}h^{\alpha\beta}\right)\left(\check\nabla_\mu+a_\mu\right)\left(\check\nabla_\nu+a_\nu\right)\Phi_{\rho\sigma}\label{eq:NRG_EOM4}\,.
\end{eqnarray}
The final equation, \eqref{eq:NRG_EOM4} above, has two typos in \cite{Hansen:2020pqs} (arXiv version 2): the last term was missing a factor of $-1/2$ in \cite{Hansen:2020pqs} and $-\left(h^{\mu\alpha}h^{\rho\beta}h^{\nu\sigma}+h^{\mu\rho}h^{\nu\beta}h^{\sigma\alpha}\right)\Phi_{\rho\sigma}\check R_{\mu\nu}$ on the 2nd-before-last line had the wrong $\alpha\beta$ indices in \cite{Hansen:2020pqs} making the expression there non-symmetric. The corrected equation is given by

Under NLO diffeomorphisms the NRG fields transform as in \eqref{eq:trafosmPhi}. If we define 
\begin{equation}
    \zeta^\mu=-\Lambda v^\mu+h^{\mu\nu}\zeta_\nu\,,
\end{equation}
then we obtain
\begin{eqnarray}
\delta h_{\mu\nu} & = & 0\,,\\
\delta m_\mu & = & \partial_\mu\Lambda -\Lambda a_\mu+h^{\rho\sigma}\zeta_\sigma a_\rho\tau_\mu\,,\\
h^{\mu\rho}h^{\nu\sigma}\delta\Phi_{\mu\nu} & = & h^{\mu\rho}h^{\nu\sigma}\left(2\Lambda K_{\mu\nu}+\check\nabla_\mu\zeta_\nu+\check\nabla_\nu\zeta_\mu\right)\,.
\end{eqnarray}
This leads to the following two Ward identities
\begin{equation}\label{eq:LambdaBianchi}
    \left(\check\nabla_\mu+2a_\mu\right)\mathcal{G}_m^\mu-K_{\mu\nu}\mathcal{G}_\Phi^{\mu\nu}=0\,,
\end{equation}
for the $\Lambda$ transformation and 
\begin{equation}\label{eq:zetaBianchi}
    \left(\check\nabla_\nu+a_\nu\right)\mathcal{G}_\Phi^{\mu\nu}-\mathcal{G}_m^\nu\tau_\nu h^{\mu\rho}a_\rho=0\,,
\end{equation}
for the $\zeta_\mu$ transformation.

Using the above expressions it can be shown that
\begin{equation}\label{eq:TTNCsourcing}
    -(D-3)\tau_\mu\mathcal{G}_m^\mu+h_{\mu\nu}\mathcal{G}_\Phi^{\mu\nu}=(D-2)h^{\mu\nu}\left(\check\nabla_\mu+a_\mu\right)a_\nu=(D-2)e^{-1}\partial_\mu\left(eh^{\mu\nu}a_\nu\right)\,.
\end{equation}
as well as
\begin{eqnarray}
&&-(D-3)\tau_\mu \mathcal{G}_{\tau}^\mu-(D-3)m_\mu \mathcal{G}_{m}^\mu+h_{\mu\nu}\mathcal{G}_{h}^{\mu\nu}+\Phi_{\mu\nu}\mathcal{G}_{\Phi}^{\mu\nu}=(D-2)\left[v^\mu v^\nu\check R_{\mu\nu}\right.\nonumber\\
&&\left.+\check\nabla_\rho\left(v^\rho h^{\mu\nu}\left(\check\nabla_\mu+a_\mu\right)m_\nu\right)+\frac{1}{4}h^{\mu\rho}h^{\nu\sigma}F_{\mu\nu}F_{\rho\sigma}+h^{\mu\nu}\left(\check\nabla_\mu+a_\mu\right)\left(\check\nabla_\nu+a_\nu\right)\phi\right.\nonumber\\
&&\left.-\left(h^{\mu\rho}h^{\nu\sigma}-\frac{1}{2}h^{\mu\nu}h^{\rho\sigma}\right)\left(\check\nabla_\mu+a_\mu\right)\left(a_\nu\Phi_{\rho\sigma}+2m_\nu K_{\rho\sigma}\right)\right]\,.\label{eq:scaleeq}
\end{eqnarray}
These expressions hold off shell.
The RHS of \eqref{eq:scaleeq} can be rewritten as
\begin{equation}
    (D-2)\left[v^\mu v^\nu\check R_{\mu\nu}+\frac{1}{4}h^{\mu\rho}h^{\nu\sigma}F_{\mu\nu}F_{\rho\sigma}+\left(\check\nabla_\rho+a_\rho\right)X^\rho\right]\,,
\end{equation}
where we defined 
\begin{equation}
    X^\rho=v^\rho h^{\mu\nu}\left(\check\nabla_\mu+a_\mu\right)m_\nu+h^{\rho\nu}\left(\check\nabla_\nu+a_\nu\right)\phi-\left(h^{\rho\kappa}h^{\nu\lambda}-\frac{1}{2}h^{\rho\nu}h^{\kappa\lambda}\right)\left(a_\nu\Phi_{\kappa\lambda}+2m_\nu K_{\kappa\lambda}\right)\,.
\end{equation}
We also used that
\begin{equation}
\left(\check\nabla_\rho+a_\rho\right)Y^\rho=e^{-1}\partial_\rho\left(eY^\rho\right)\,,
\end{equation}
for any $Y^\rho$.
We can then rewrite $X^\rho$ as
\begin{equation}
    X^\rho=h^{\rho\sigma}v^\alpha F_{\alpha\sigma}+v^\rho h^{\mu\nu}a_\mu m_\nu+h^{\rho\nu}a_\nu I-h^{\rho\kappa}h^{\nu\lambda}a_\nu\Phi_{\kappa\lambda}+\check\nabla_\mu A^{\rho\mu}\,,
\end{equation}
where we defined the antisymmetric tensor
\begin{equation}
    A^{\rho\mu}=v^\rho h^{\mu\nu}m_\nu-v^\mu h^{\rho\nu}m_\nu\,,
\end{equation}
for which it can be shown that
\begin{equation}
    \check\nabla_\mu A^{\rho\mu}=e^{-1}\partial_\mu\left(e A^{\rho\mu}\right)\,.
\end{equation}
It follows that 
\begin{equation}
    \left(\check\nabla_\rho+a_\rho\right)\check\nabla_\mu A^{\rho\mu}=0\,.
\end{equation}
So in the end we find that \eqref{eq:scaleeq} can be written as
\begin{align}
    &-(D-3)\tau_\mu \mathcal{G}_{\tau}^\mu-(D-3)m_\mu \mathcal{G}_{m}^\mu+h_{\mu\nu}\mathcal{G}_{h}^{\mu\nu}+\Phi_{\mu\nu}\mathcal{G}_{\Phi}^{\mu\nu} \nonumber
    \\
    &=(D-2)\left[v^\mu v^\nu\check R_{\mu\nu}+\frac{1}{4}h^{\mu\rho}h^{\nu\sigma}F_{\mu\nu}F_{\rho\sigma}+e^{-1}\partial_\rho\left(e\tilde X^\rho\right)\right]\,, \label{eq:SimplEOM2}
\end{align}
where 
\begin{align}
    \tilde X^\rho & =  h^{\rho\sigma}v^\mu F_{\mu\sigma}+v^\rho h^{\mu\nu}a_\mu m_\nu+h^{\rho\nu}a_\nu I-h^{\rho\kappa}h^{\nu\lambda}a_\nu\Phi_{\kappa\lambda}\nonumber\\
    & =  h^{\rho\sigma}v^\mu\left(\partial_\mu m_\sigma-\partial_\sigma m_\mu\right)+v^\rho h^{\mu\nu}a_\mu m_\nu-\frac{1}{2}h^{\rho\kappa}h^{\sigma\lambda}\left(a_\sigma\Phi_{\lambda\kappa}+a_\lambda\Phi_{\sigma\kappa}-a_\kappa\Phi_{\sigma\lambda}\right)\,.
\end{align}
We will use both \eqref{eq:SimplEOM2} and \eqref{eq:TTNCsourcing} again in section \ref{app:Simplification} when we write down the simplified version of the NRG EOM sourced by only a negative cosmological constant.

\subsection{Comparing with the $1/c^2$ expansion of Einstein's equations} \label{app:ComparingWithOn-shell}
We next show that the NRG EOM can also be obtained from a direct $1/c^2$ expansion of the Einstein's equations. Borrowing from \cite{Hartong:2023ckn} we know that the Einstein equations in PNR variables are
\begin{align}\label{eq:PNRdecomRicci}
    R_{\mu \nu} = c^4 \RA_{\mu \nu} + c^2 \RB_{\mu \nu}+  \RC_{\mu \nu}+ c^{-2} \RD_{\mu \nu}=\frac{2}{D-2}\Lambda\left(-c^2 T_\mu T_\nu+\Pi_{\mu\nu}\right)\,,
\end{align}
where $D$ is the number of spacetime dimensions and where we assumed that there is only a cosmological constant and no matter. We also used the definitions 
\begin{align}
\RA_{\mu \nu} &= \frac{1}{4} T_\mu T_\nu \Pi^{\alpha \beta} \Pi^{\rho \sigma} T_{\alpha \rho} T_{\beta \sigma}\,, \label{ER-4}\\
\RB_{\mu \nu} &= \overset{(C)}{\nabla}_\sigma {W^\sigma_{\mu \nu}} +{W^\sigma_{\mu \nu}} S^\lambda_{\lambda \sigma} - {W^\sigma_{\mu \lambda}} S^\lambda_{\sigma \nu} -{W^\sigma_{\nu\lambda }} S^\lambda_{\sigma\mu } \,,\label{ER-2}
\\
\RC_{\mu \nu} &= \overset{(C)}{R}_{\mu \nu} - {W^{\sigma}_{\mu \lambda}} {V^{\lambda}_{\sigma \nu}} - {W^{\sigma}_{\nu \lambda}} {V^{\lambda}_{\sigma \mu}}  - \overset{(C)}{\nabla}_{\mu}  S^{\sigma}_{\sigma \nu} + \overset{(C)}{\nabla}_{\sigma} S^{\sigma}_{\mu \nu} -2C^{\lambda}_{[\mu \sigma]}S^{\sigma}_{\lambda \nu}\,, \label{check}
\\
\RD_{\mu \nu} &= \overset{(C)}{\nabla}_\sigma {V^\sigma_{\mu \nu}}\,, \label{ER2}
\end{align}
where 
\begin{align}
{W^\rho_{\mu \nu}} := & \frac{1}{2} T_\mu \Pi^{\rho \sigma} (\partial_\sigma T_\nu - \partial_\nu T_\sigma) + \frac{1}{2}T_\nu \Pi^{\rho \sigma} (\partial_\sigma T_\mu - \partial_\mu T_\sigma)\,, \label{C(-2)}\\
C^\rho_{\mu \nu}  :=& - T^\rho \partial_\mu T_\nu + \frac{1}{2} \Pi^{\rho \sigma} (\partial_{\nu} \Pi_{\mu \sigma} + \partial_{\mu} \Pi_{\nu\sigma} - \partial_{\sigma} \Pi_{\mu \nu})\,,\label{C-conn}\\
S^\rho_{\mu \nu} :=& \frac{1}{2} T^\rho (\partial_\mu T_\nu - \partial_\nu T_\mu - T_\mu \sL_T T_\nu - T_\nu \sL_T T_\mu)\,, \label{C(0)}\\
{V^\rho_{\mu \nu}} := & \frac{1}{2} T^\rho \mathcal{L}_{T} \Pi_{\mu \nu}\,.\label{C(2)}
\end{align}
In here $\mathcal{L}_T$ denotes the Lie derivative along $T^\mu$. The object \eqref{C-conn} plays the role of a connection in the PNR formulation of GR. A superscript $(C)$ indicates that the connection \eqref{C-conn} has been used.

It can be shown that
\begin{eqnarray}
    \RB_{\mu \nu} & = & T_\mu T_\nu \Pi^{\rho\sigma}\overset{(C)}{\nabla}_\rho\mathcal{L}_T T_\sigma+T_\mu T_\nu \Pi^{\rho\sigma}\mathcal{L}_T T_\rho \mathcal{L}_T T_\sigma+T_{(\mu}\Pi_{\nu)\sigma}\overset{(C)}{\nabla}_\rho T^{\rho\sigma}_\Pi\nonumber\\
    &&+2T_{(\mu}\Pi_{\nu)\sigma}\mathcal{L}_T T_\rho T^{\rho\sigma}_\Pi+\frac{1}{2}\Pi_{\mu\rho}\Pi_{\nu\sigma}\Pi_{\kappa\lambda}T^{\rho\kappa}_\Pi T^{\sigma\lambda}_\Pi\,,\\
    \RC_{\mu \nu} & = & \overset{(C)}{R}_{\mu\nu}+T_{(\mu}\mathcal{L}_T\Pi_{\nu)\rho}\Pi^{\rho\sigma}\mathcal{L}_T T_\sigma-\Pi_{\mu\sigma}\Pi^{\rho\sigma}\overset{(C)}{\nabla}_\rho\left(\mathcal{L}_T T_\nu\right)-\mathcal{L}_T T_\mu \mathcal{L}_T T_\nu\nonumber\\
    &&-\frac{1}{2}\Pi^{\rho\sigma}\mathcal{L}_T\Pi_{\rho\sigma}T_\mu\mathcal{L}_T T_\nu-\frac{1}{2}\Pi_{\rho(\mu}\mathcal{L}_T T_{\nu)\sigma}T_\Pi^{\rho\sigma}\nonumber\\
    &&+\frac{1}{4}\Pi^{\rho\sigma}\mathcal{L}_T \Pi_{\rho\sigma}\Pi_{\mu\alpha}\Pi_{\nu\beta}T^{\alpha\beta}_\Pi+\frac{1}{2}\Pi_{\mu\alpha}\Pi_{\nu\beta}T^\rho\overset{(C)}{\nabla}_\rho T^{\alpha\beta}_\Pi\,,\nonumber\\
    & = & \overset{(C)}{R}_{(\mu\nu)}+T_{(\mu}\mathcal{L}_T\Pi_{\nu)\rho}\Pi^{\rho\sigma}\mathcal{L}_T T_\sigma-\Pi^{\rho\sigma}\Pi_{\sigma(\mu}\overset{(C)}\nabla_{|\rho|}\mathcal{L}_T T_{\nu)}-\mathcal{L}_T T_\mu\mathcal{L}_T T_\nu\nonumber\\
    &&-\frac{1}{2}\Pi^{\rho\sigma}\mathcal{L}_T\Pi_{\rho\sigma}T_{(\mu}\mathcal{L}_T T_{\nu)}-\frac{1}{2}\Pi_{\rho(\mu}\mathcal{L}_T \Pi_{\nu)\sigma} T^{\rho\sigma}_\Pi\,,
\end{eqnarray}
where we defined
\begin{equation}
    T^{\rho\sigma}_\Pi=\Pi^{\rho\alpha}\Pi^{\sigma\beta}T_{\alpha\beta}\,,
\end{equation}
and where we used
\begin{equation}
    \overset{(C)}{\nabla}_\mu T^\nu=\frac{1}{2}\Pi^{\nu\rho}\mathcal{L}_T\Pi_{\rho\mu}\,,\qquad \overset{(C)}{\nabla}_\mu\Pi_{\nu\rho}=T_{(\nu}\mathcal{L}_T\Pi_{\rho)\mu}\,,
\end{equation}
as well as the expression for $\overset{(C)}{R}_{[\mu\nu]}$ given in equation \eqref{eq:ASpartRicci} in appendix \ref{app:curvature}.
Expanding \eqref{eq:PNRdecomRicci} we find
\begin{align}
     \RA_{\mu \nu}\Big|_{\text{LO}} & =  0\,, \nonumber\\
    \RA_{\mu \nu}\Big|_{\text{NLO}} +  \RB_{\mu \nu}\Big|_{\text{LO}} & =  -\frac{2}{D-2}\Lambda T_\mu T_\nu\Big|_{\text{LO}}\,, \nonumber \\
    \RA_{\mu \nu}\Big|_{\text{NNLO}} + \RB_{\mu \nu}\Big|_{\text{NLO}}+  \RC_{\mu \nu}\Big|_{\text{LO}} & =  -\frac{2}{D-2}\Lambda T_\mu T_\nu\Big|_{\text{NLO}}+\frac{2}{D-2}\Lambda\Pi_{\mu\nu}\Big|_{\text{LO}}\,, \nonumber\\
    \RA_{\mu \nu}\Big|_{\text{N$^3$LO}} + \RB_{\mu \nu}\Big|_{\text{NNLO}}+  \RC_{\mu \nu}\Big|_{\text{NLO}}+  \RD_{\mu \nu}\Big|_{\text{LO}} & =  -\frac{2}{D-2}\Lambda T_\mu T_\nu\Big|_{\text{NNLO}}+\frac{2}{D-2}\Lambda\Pi_{\mu\nu}\Big|_{\text{NLO}}\,.\label{eq:NNLO-EOM}
\end{align}

The first of these equations is 
\begin{equation}
    \RA_{\mu \nu}\Big|_{\text{LO}}=\frac{1}{4} \tau_\mu \tau_\nu h^{\alpha \beta} h^{\rho \sigma} \tau_{\alpha \rho} \tau_{\beta \sigma}=0\,,
\end{equation}
which is the familiar TTNC condition. We will use this in simplifying the subsequent equations. In particular using the TTNC condition we see that 
\begin{align}
    \RA_{\mu \nu}\Big|_{\text{NLO}} & =  0\,,\\
    \RA_{\mu \nu}\Big|_{\text{NNLO}} & =  \frac{1}{4}\tau_\mu\tau_\nu h^{\alpha\beta}h^{\rho\sigma}F_{\alpha\rho}F_{\beta\sigma}\,,\\
    \RA_{\mu \nu}\Big|_{\text{NNNLO}} & =  \frac{1}{4}\tau_\mu\tau_\nu\left(\Pi^{\alpha\beta}\Pi^{\rho\sigma}T_{\alpha\rho}T_{\beta\sigma}\right)\Big|_{\text{NNNLO}}+\frac{1}{4}\left(m_\mu\tau_\nu+m_\nu\tau_\mu\right)h^{\alpha\beta}h^{\rho\sigma}F_{\alpha\rho}F_{\beta\sigma}\,.
\end{align}
Furthermore, we have
\begin{eqnarray}
    \RB_{\mu \nu}\Big|_{\text{LO}} & = & \tau_\mu\tau_\nu\left(h^{\rho\sigma}\check\nabla_\rho a_\sigma+h^{\rho\sigma}a_\rho a_\sigma\right)\,,\\
    \RB_{\mu \nu}\Big|_{\text{NLO}} & = & \tau_\mu\tau_\nu\left(e^{-1}\partial_\rho\left[eh^{\rho\sigma}v^\alpha F_{\alpha\sigma}+ev^\rho h^{\sigma\alpha}a_\sigma m_\alpha-eh^{\rho\alpha}h^{\sigma\beta}\Phi_{\alpha\beta}a_\sigma\right]+h^{\rho\sigma}\partial_\rho I a_\sigma\right)\nonumber\\
    &&+2\tau_{(\mu}m_{\nu)}\left(h^{\rho\sigma}\check\nabla_\rho a_\sigma+h^{\rho\sigma}a_\rho a_\sigma\right)+\tau_{(\mu}h_{\nu)\sigma}h^{\rho\alpha}h^{\sigma\beta}\left(\check\nabla_\rho+2a_\rho\right)F_{\alpha\beta}\,,
\end{eqnarray}
and
\begin{eqnarray}
    \RC_{\mu \nu}\Big|_{\text{LO}} & = & \check R_{(\mu\nu)}+\tau_{(\mu}\mathcal{L}_v h_{\nu)\rho}h^{\rho\sigma}a_\sigma-h^{\rho\sigma}h_{\sigma(\mu}\check\nabla_{|\rho|}a_{\nu)}-a_\mu a_\nu-\frac{1}{2}h^{\rho\sigma}\mathcal{L}_v h_{\rho\sigma}\tau_{(\mu}a_{\nu)}\nonumber\\
    & = & \check R_{\mu\nu}-P_\mu^\rho P_\nu^\sigma\left(\check\nabla_\rho+a_\rho\right)a_\sigma-\tau_\mu a_\rho h^{\rho\sigma}K_{\nu\sigma}+\tau_\mu K a_\nu\,,
\end{eqnarray}
where we used 
\begin{equation}\label{eq:ASRic}
    \check R_{[\mu\nu]} = -\frac{1}{2}K\left(\tau_\mu a_\nu-\tau_\nu a_\mu\right)+\frac{1}{2}v^\rho\tau_\mu\check\nabla_\nu a_\rho-\frac{1}{2}v^\rho\tau_\nu\check\nabla_\mu a_\rho\,.
\end{equation}
The last equation in \eqref{eq:NNLO-EOM} contains NNLO fields except when we contract it with $h^{\mu\kappa} h^{\nu\lambda}$ in which case we obtain
\begin{equation}
    h^{\mu\kappa} h^{\nu\lambda}\left(
     \RB_{\mu \nu}\Big|_{\text{NNLO}}+  \RC_{\mu \nu}\Big|_{\text{NLO}}+  \RD_{\mu \nu}\Big|_{\text{LO}} \right)= h^{\mu\kappa} h^{\nu\lambda}\left( -\frac{2}{D-2}\Lambda m_\mu m_\nu+\frac{2}{D-2}\Lambda\Phi_{\mu\nu}\right)\,,
\end{equation}
where 
\begin{eqnarray}
    h^{\mu\kappa} h^{\nu\lambda}\RB_{\mu \nu}\Big|_{\text{NNLO}} & = &  h^{\mu\kappa} h^{\nu\lambda}m_\mu m_\nu h^{\rho\sigma}\left(\check\nabla_\rho+a_\rho\right)a_\sigma+h^{\mu(\kappa}m_\mu h^{\lambda)\sigma}\left(\check\nabla_\rho+2a_\rho\right)h^{\rho\nu}F_{\nu\sigma}\nonumber\\
    &&+\frac{1}{2}h^{\kappa\mu}h^{\lambda\nu}h^{\rho\sigma}F_{\mu\rho}F_{\nu\sigma}\,,\\
    h^{\mu\kappa} h^{\nu\lambda}\RC_{\mu \nu}\Big|_{\text{NLO}} & = & h^{\mu\kappa} h^{\nu\lambda}\left( U_{\mu\alpha\nu}{}^\alpha+\bar\chi^\rho_{\mu\nu}a_\rho-2h^{\rho\sigma}a_\sigma m_{(\mu}K_{\nu)\rho}-h^{\rho\sigma}a_\sigma m_{[\mu}K_{\nu]\rho}+Km_\mu a_\nu\right.\nonumber\\
    &&\left.-2v^\alpha F_{\alpha(\mu}a_{\nu)}-m_{(\mu}v^\rho\check\nabla_{|\rho|}a_{\nu)}+K_{\mu\nu}h^{\rho\sigma}a_\rho m_\sigma-h^{\rho\sigma}K_{\rho[\mu}F_{\nu]\sigma}\right.\nonumber\\
    &&\left.-\frac{1}{2}K F_{\mu\nu}+v^\alpha\check\nabla_{(\mu}F_{\nu)\alpha}\right)\,,\label{eq:R0hhNLO}\\
    h^{\mu\kappa} h^{\nu\lambda}\RD_{\mu \nu}\Big|_{\text{LO}} & = & h^{\mu\kappa} h^{\nu\lambda}\left(K K_{\mu\nu}-v^\rho\check\nabla_\rho K_{\mu\nu}\right)\,,
\end{eqnarray}
and where we defined
\begin{equation}
    \bar\chi^\rho_{\mu\nu}=\frac{1}{2}h^{\rho\sigma}\left(\check\nabla_\mu\Phi_{\nu\sigma}+\check\nabla_\nu\Phi_{\mu\sigma}-\check\nabla_\sigma\Phi_{\mu\nu}\right)\,.
\end{equation}

The $1/c^2$ expansion of the Einstein equations to the orders worked out above are equivalent to the EOM obtained by varying the NRG Lagrangian as we show next. With a bit of work one can show that the following identities hold assuming only the TTNC condition:
\begin{eqnarray}
   \hspace{-1cm} (D-2)v^\mu v^\nu\left(\RA_{\mu \nu}\Big|_{\text{NLO}} +  \RB_{\mu \nu}\Big|_{\text{LO}}\right) &=&-(D-3)\tau_\mu\mathcal{G}_m^\mu+h_{\mu\nu}\mathcal{G}_\Phi^{\mu\nu}\,,\\
   \hspace{-1cm}(D-2)v^\mu v^\nu\left(\RA_{\mu \nu}\Big|_{\text{NNLO}} + \RB_{\mu \nu}\Big|_{\text{NLO}}+  \RC_{\mu \nu}\Big|_{\text{LO}}\right) &= &h_{\mu\nu}\mathcal{G}^{\mu\nu}_h+\Phi_{\mu\nu}\mathcal{G}_\Phi^{\mu\nu}\label{eq:vvNLOEOM}\\
 \hspace{-1cm}  &&-(D-3)\left(\tau_\mu\mathcal{G}_\tau^\mu+m_\mu\mathcal{G}^\mu_m\right)\,,\nonumber\\
  \hspace{-1cm} h^{\mu\alpha}v^\nu\left(\RA_{\mu \nu}\Big|_{\text{NNLO}} + \RB_{\mu \nu}\Big|_{\text{NLO}}+  \RC_{\mu \nu}\Big|_{\text{LO}}\right)&=&-P^\alpha_\mu\mathcal{G}^\mu_m\,,\\
  \hspace{-1cm} \left(h^{\mu\alpha}h^{\nu\beta}-h^{\alpha\beta}h^{\mu\nu}\right)\left(\RA_{\mu \nu}\Big|_{\text{NNLO}} + \RB_{\mu \nu}\Big|_{\text{NLO}}+  \RC_{\mu \nu}\Big|_{\text{LO}}\right)&=&\mathcal{G}_\Phi^{\alpha\beta}+h^{\alpha\beta}\tau_\mu\mathcal{G}^\mu_m\,. \label{eq:PhiEOM.}
\end{eqnarray}
\begin{eqnarray}
    \hspace{-1cm}&&\left(h^{\mu\alpha}h^{\nu\beta}-h^{\mu\nu}h^{\alpha\beta}\right)\left(\RB_{\mu \nu}\Big|_{\text{NNLO}}+  \RC_{\mu \nu}\Big|_{\text{NLO}}+  \RD_{\mu \nu}\Big|_{\text{LO}}\right)=\mathcal{G}_h^{\mu\nu}P_\mu^\alpha P_\nu^\beta+h^{\alpha\beta}\tau_\mu\mathcal{G}^\mu_\tau\nonumber\\
   \hspace{-1cm} &&+2h^{\mu(\alpha}m_\mu P_\nu^{\beta)}\mathcal{G}_m^\nu-h^{\alpha\beta}m_\mu P^\mu_\nu\mathcal{G}^\nu_m-\phi\mathcal{G}^{\alpha\beta}_\Phi+\frac{1}{2}h^{\alpha\beta}h^{\rho\sigma}\Phi_{\rho\sigma}\tau_\mu \mathcal{G}^\mu_m-h^{\mu\alpha}h^{\nu\beta}\Phi_{\mu\nu}\tau_\rho\mathcal{G}^\rho_m\nonumber\\
    \hspace{-1cm}&&-\frac{1}{2}h^{\rho\sigma}\Phi_{\rho\sigma}\mathcal{G}_\Phi^{\alpha\beta}-h^{\alpha\beta}\Phi_{\mu\nu}\mathcal{G}^{\mu\nu}_\Phi+2h^{\rho(\alpha}\Phi_{\rho\sigma}\mathcal{G}_\Phi^{\beta)\sigma}\nonumber\\
   \hspace{-1cm} &&-\left(h^{\mu\alpha}h^{\nu\beta}-h^{\mu\nu}h^{\alpha\beta}\right)\left(\Phi_{\mu\nu}-m_\mu m_\nu\right)h^{\rho\sigma}\left(\check\nabla_\rho+a_\rho\right)a_\sigma\,. \label{eq:EOM-Simplification}
\end{eqnarray}
Thus, we see that the expanded Einstein field equations follow from the NRG EOM. So now we just need to show that these relations are invertible. To do that we see that taking the trace of the equation \eqref{eq:EOM-Simplification} we find
\begin{eqnarray}
    &&-(D-2)h^{\mu\nu}\left[\RB_{\mu \nu}\Big|_{\text{NNLO}}+  \RC_{\mu \nu}\Big|_{\text{NLO}}+  \RD_{\mu \nu}\Big|_{\text{LO}}\right]=\nonumber\\
    &&h_{\mu\nu}\mathcal{G}_h^{\mu\nu}+(D-1)\tau_\mu\mathcal{G}^\mu_\tau-(D-3)m_\mu \mathcal{G}^\mu_m-(D-3)\Phi_{\mu\nu}\mathcal{G}^{\mu\nu}_\Phi\nonumber\\
    &&+(D-2)\left(h^{\rho\sigma}\left(\Phi_{\rho\sigma}-m_\rho m_\sigma\right)-I\right)h^{\mu\nu}\left(\check\nabla_\mu+a_\mu\right)a_\nu\,.\label{eq:tracehhNNLOEOM}
\end{eqnarray}
If we solve \eqref{eq:vvNLOEOM} for $h_{\mu\nu}\mathcal{G}_h^{\mu\nu}-(D-3)m_\mu \mathcal{G}^\mu_m$ and \eqref{eq:PhiEOM.} for $\mathcal{G}_\Phi^{\mu \nu}$, we can substitute the result in \eqref{eq:tracehhNNLOEOM} and obtain an expression for $\tau_\mu\mathcal{G}^\mu_\tau$ for $D\ge 3$. From there it is easy to find expressions for all the rest of the NRG EOM. We thus see that for $D\ge 3$ there is a one-to-one relation between the responses to varying the NRG Lagrangian and the expansion of the Einstein equations up to the order specified above. The case $D=2$ as usual requires special treatment and we have not looked into this case.

\subsection{Simplifying the EOM} \label{app:Simplification}
We add a cosmological constant (and no matter source) to the NRG Lagrangian in the form of \eqref{eq:LambdaAction}. Varying the cosmological constant term we then obtain the sources and end up with the following equations of motion
\begin{eqnarray}
e^{-1}\partial_\mu\left(eh^{\mu\nu}a_\nu\right) & = & -\frac{2}{D-2}\Lambda\,,\label{eq:diva-app}\\
    e^{-1}\partial_\mu\left(e\tilde X^\mu\right)+v^\mu v^\nu\check R_{\mu\nu}+\frac{1}{4} h^{\mu\nu}h^{\rho\sigma}F_{\mu\rho}F_{\nu\sigma}
    & = & -\frac{2}{D-2}\Lambda I\,, \label{eq:divX-app}\\
-2h^{\nu\sigma}v^\alpha\check R_{\nu\alpha}+\left(\check\nabla_\rho+2a_\rho\right)h^{\rho\mu}h^{\sigma\nu}F_{\mu\nu} & = & 0\,,\label{eq:PvNLOEOM-app}\\
h^{\mu\kappa} h^{\nu\lambda}\left(\check R_{\mu\nu}-\check\nabla_\mu a_{\nu}-a_\mu a_\nu\right) & = & \frac{2}{D-2}\Lambda h^{\kappa\lambda}\,,\label{eq:NLO-EOMhh-app}
\end{eqnarray}
where \eqref{eq:diva-app} and \eqref{eq:divX-app} follow from \eqref{eq:SimplEOM2} and \eqref{eq:TTNCsourcing} respectively. Meanwhile, \eqref{eq:PvNLOEOM-app} and \eqref{eq:NLO-EOMhh-app} follows from rewriting \eqref{eq:NRG_EOM2} and \eqref{eq:NRG_EOM3}.
Regarding equation \eqref{eq:PvNLOEOM-app} we notice that we have the identity
\begin{equation}
2h^{\nu\sigma}h^{\mu\rho}\left(\check\nabla_{\mu}K_{\nu\rho}-\check\nabla_{\nu}K_{\mu\rho}\right)=-2h^{\nu\sigma}v^\alpha\check R_{\nu\alpha}\,.
\end{equation}
Taking the trace of \eqref{eq:NLO-EOMhh-app} and using \eqref{eq:diva-app} leads to
\begin{equation}
    h^{\mu\nu}\check R_{\mu\nu}=2\Lambda\,,
\end{equation}
which is the equation of motion of $\phi=-v^\mu m_\mu$ (equation \eqref{eq:Phi-EOM}). Finally, instead of the long equation \eqref{eq:NRG_EOM4} we use what we learned from \eqref{eq:EOM-Simplification} to rewrite it as
\begin{eqnarray}
    &&h^{\mu\kappa} h^{\nu\lambda}\left(m_\mu v^\rho\left(\check R_{(\nu\rho)}-\check\nabla_{(\nu}a_{\rho)}\right)+m_\nu v^\rho\left(\check R_{(\mu\rho)}-\check\nabla_{(\mu}a_{\rho)}\right)-  U_{\alpha(\mu\nu)}{}^\alpha+\chi^\rho_{\mu\nu}a_\rho\right.\\
    &&\left.+v^\rho\left(\check\nabla_{(\mu}+2a_{(\mu}\right)F_{\nu)\rho}+K K_{\mu\nu}-v^\rho\check\nabla_\rho K_{\mu\nu}+\frac{1}{2}h^{\rho\sigma}F_{\mu\rho}F_{\nu\sigma}\right)=\frac{2}{D-2}\Lambda h^{\mu\kappa} h^{\nu\lambda} \Phi_{\mu\nu}\,,\nonumber
\end{eqnarray}
where in writing it in this form we used equations \eqref{eq:diva-app} and \eqref{eq:PvNLOEOM-app} as well as the fact that the antisymmetric part of \eqref{eq:R0hhNLO} is zero, and finally also the identity
\begin{equation}
    v^\rho\check R_{\nu\rho}+\frac{1}{2}Ka_{\nu}-h^{\rho\sigma}a_\sigma K_{\nu\rho}-\frac{1}{2}v^\rho\check\nabla_{\rho}a_{\nu}=v^\rho\left(\check R_{(\nu\rho)}-\check\nabla_{(\nu}a_{\rho)}\right)\,,
\end{equation}
which follows from \eqref{eq:ASRic}.

\section{Ansatz for solutions with planar symmetries}\label{app:ansatz}

In this appendix, we assume that the LO geometry is given by equation \eqref{eq:NRAdS2} and we want to find the most general $\Phi_{ab}$\footnote{We will split the coordinates of the LO geometry as $x^\mu=(t,x^a)$ where $x^a=(r, x^i)$.} and $m_\mu$ (up to a gauge transformation) that respect time translation, boundary space translation and boundary rotation symmetries. We do this specifically for $d=3$ but we expect the argument to hold for general dimensions. 

\subsection{Gauge invariant curvatures}
We start by constructing a gauge invariant curvature for $\Phi_{\mu\nu}$. We define 
\begin{equation}
    U_{\mu\nu\rho\sigma}=U_{\mu\nu\rho}{}^\lambda h_{\lambda\sigma}+\check R_{\mu\nu\rho}{}^\lambda\Phi_{\lambda\sigma}\,.
\end{equation}
This transforms under NLO diffeomorphisms as
\begin{equation}
    \delta_\zeta U_{\mu\nu\rho\sigma}=\mathcal{L}_\zeta\left(\check R_{\mu\nu\rho}{}^\lambda h_{\lambda\sigma}\right)\,.
\end{equation}
Since we assume the leading order geometry is given by \eqref{eq:NRAdS2} we can use equation \eqref{eq:vacRiem} to see that
\begin{equation}
    \delta_\zeta U_{\mu\nu\rho\sigma}=-l^{-2}\mathcal{L}_\zeta\left(h_{\mu\rho}h_{\nu\sigma}-h_{\mu\sigma}h_{\nu\rho}\right)\,.
\end{equation}
We then define
\begin{equation}
    \tilde U_{\mu\nu\rho\sigma}=U_{\mu\nu\rho\sigma}+h^{\kappa\lambda}a_\kappa a_\lambda\left(h_{\mu\rho}\Phi_{\nu\sigma}+h_{\nu\sigma}\Phi_{\mu\rho}-h_{\nu\rho}\Phi_{\mu\sigma}-h_{\mu\sigma}\Phi_{\nu\rho}\right)\,,
\end{equation}
where $h^{\kappa\lambda}a_\kappa a_\lambda=l^{-2}$. The expression for $\tilde U_{\mu\nu\rho\sigma}$ is invariant under NLO diffeomorphisms ($\zeta^\mu$).

Only the spatial projection of $\Phi_{\mu\nu}$ appears in NRG whose curvature is the spatial projection of $U_{\mu\nu\rho\sigma}$ or $\tilde U_{\mu\nu\rho\sigma}$. With that in mind, we see that
\begin{eqnarray}
    \hspace{-.5cm}&&h^{\mu\alpha}h^{\nu\beta}h^{\rho\gamma}h^{\sigma\delta} U_{\mu\nu\rho\sigma}  = h^{\mu\alpha}h^{\nu\beta}h^{\rho\gamma}h^{\sigma\delta} \left( -\check\nabla_\mu\bar\chi_{\nu\rho\sigma}+\check\nabla_\nu\bar\chi_{\mu\rho\sigma}+\check R_{\mu\nu\rho}{}^\lambda\Phi_{\lambda\sigma}\right.\nonumber\\
    \hspace{-.5cm}&&\left.-m_\sigma\left(\check\nabla_\mu K_{\nu\rho}-\check\nabla_\nu K_{\mu\rho}\right)-K_{\nu\rho}\check\nabla_\mu m_\sigma+K_{\mu\rho}\check\nabla_\nu m_\sigma-K_{\mu\sigma}\check\nabla_\nu m_\rho+K_{\nu\sigma}\check\nabla_\mu m_\rho\right)\,,
\end{eqnarray}
where 
\begin{equation}
    \bar\chi_{\nu\rho\sigma}=\frac{1}{2}\left(\check\nabla_\nu\Phi_{\rho\sigma}+\check\nabla_\rho\Phi_{\nu\sigma}-\check\nabla_\sigma\Phi_{\nu\rho}\right)\,.
\end{equation}
When projected with $h^{\mu\alpha}h^{\nu\beta}h^{\rho\gamma}h^{\sigma\delta}$ the tensors $U_{\mu\nu\rho\sigma}$ and $\tilde U_{\mu\nu\rho\sigma}$ both have the same symmetry properties as the $(0,4)$ Riemann tensor in GR, i.e. 
\begin{equation}\label{eq:indexprop}
    U_{\mu\nu\rho\sigma}=-U_{\nu\mu\rho\sigma}=-U_{\mu\nu\sigma\rho}=U_{\rho\sigma\mu\nu}\,,\qquad U_{[\mu\nu\rho\sigma]}=0\,,
\end{equation}
and similarly for $\tilde U_{\mu\nu\rho\sigma}$ where we assumed that all indices are spatially projected.

We also have a field strength for $m_\mu$ which is given by $F_{\mu\nu}$. Equation \eqref{eq:trafohhF} tells us that $h^{\mu\rho}h^{\nu\sigma} \delta_\zeta F_{\mu\nu}$ is invariant under NLO diffeomorphisms for any LO geometry that obeys TTNC.

We thus have two $\zeta^\mu$ gauge invariant field strengths, $\tilde U_{abcd}$ and $F_{ab}$. When we want to make an ansatz for $m_\mu$ and the spatial part of $\Phi_{\mu\nu}$ that respects certain Killing symmetries then their field strengths must obey
\begin{equation}\label{eq:Killingsym}
    h^{\mu\alpha}h^{\nu\beta}h^{\rho\gamma}h^{\sigma\delta}\mathcal{L}_K\tilde U_{\mu\nu\rho\sigma}=0\,,\qquad h^{\mu\alpha}h^{\nu\beta}\mathcal{L}_K F_{\mu\nu}=0\,,
\end{equation}
where $K^\mu$ is one of the Killing vectors of the desired symmetry. We can then work out what this means for the gauge fields $m_\mu$ and $\Phi_{\mu\nu}$ in a certain gauge.

\subsection{Imposing planar symmetry}
The total gauge transformation acting on $m_\mu$ and $h^{\mu\alpha}h^{\nu\beta}\Phi_{\mu\nu}$ is
\begin{eqnarray}
    \delta m_\mu & = & \mathcal{L}_\xi m_\mu+\mathcal{L}_\zeta\tau_\mu+\lambda_\mu\,,\\
    \delta \Phi_{\mu\nu} & = & \mathcal{L}_\xi \Phi_{\mu\nu}+\mathcal{L}_\zeta h_{\mu\nu}+\lambda_\mu m_\nu+\lambda_\nu m_\mu+\kappa_\mu\tau_\nu+\kappa_\nu\tau_\mu\,.
\end{eqnarray}
A symmetry is a diffeomorphism generated by $K^\mu$ such that 
\begin{eqnarray}
    0=\delta m_\mu & = & \mathcal{L}_K m_\mu+\mathcal{L}_\zeta\tau_\mu+\lambda_\mu\,,\\
    0=\delta \Phi_{\mu\nu} & = & \mathcal{L}_K \Phi_{\mu\nu}+\mathcal{L}_\zeta h_{\mu\nu}+\lambda_\mu m_\nu+\lambda_\nu m_\mu+\kappa_\mu\tau_\nu+\kappa_\nu\tau_\mu\,,
\end{eqnarray}
for appropriate choices of $\zeta^\mu$, $\lambda_\mu$ and $\kappa_\mu$.

The Galilean boosts are fixed by setting $h_{ti}=0$ and so we will not consider these any further. We then have
\begin{eqnarray}
    0=\delta \Phi_{\mu\nu} & = & \mathcal{L}_K \Phi_{\mu\nu}+\mathcal{L}_\zeta h_{\mu\nu}+\kappa_\mu\tau_\nu+\kappa_\nu\tau_\mu\,.
\end{eqnarray}
The subleading boosts with parameter $\kappa_\mu$ are irrelevant for the spatial components of $\Phi_{\mu\nu}$ since the latter transform as
\begin{eqnarray}
    \delta\Phi_{rr} & = & 2\partial_r\zeta_r+2r^{-1}\zeta_r\,,\\
    \delta\Phi_{ir} & = & \partial_i\zeta_r+\partial_r\zeta_i+2r^{-1}\zeta_i\,,\\
    \delta\Phi_{ij} & = & \partial_i\zeta_j+\partial_j\zeta_i-2r^{-1}\delta_{ij}\zeta_r\,,\label{eq:trafoPhiij}
\end{eqnarray}
where $\zeta_\mu=h_{\mu\nu}\zeta^\nu$ and where we used the LO geometry is given by \eqref{eq:NRAdS2}.

We would like to find the most general $m_t$, $m_a$ and $\Phi_{ab}$ (up to a gauge transformation) that are compatible with the following Killing symmetries
\begin{equation}\label{eq:Killingvec}
    K=\partial_t\,,\partial_i\,,x^i\partial_j-x^j\partial_i\,,
\end{equation}
i.e. boundary time and space translation and boundary rotation invariance, where $i,j,k=1,2,3$.
We will implement the Killing symmetries by demanding that $\tilde U_{abcd}$ and $F_{ab}$ are invariant under these transformations.

\subsubsection{Constructing planar symmetric $\Phi_{ab}$}
We start with $\tilde U_{abcd}$. It is easy to see that the above Killing vectors tell us that we must have
\begin{eqnarray}
    \tilde U_{ijkl} & = & E(r)\left(\delta_{ik}\delta_{jl}-\delta_{il}\delta_{jk}\right)\,,\label{eq:tildeUijkl}\\
    \tilde U_{rijk} & = & 0\,,\label{eq:tildeUijk}\\
    \tilde U_{rirj} & = & G(r)\delta_{ij}\,,\label{eq:tildeUij}
\end{eqnarray}
where we used \eqref{eq:indexprop}, \eqref{eq:Killingsym} and \eqref{eq:Killingvec}, as well as the fact that the only $SO(3)$ invariant tensors are $\delta_{ij}$ and $\epsilon_{ijk}$. The properties \eqref{eq:indexprop} imply that there can be no epsilon tensor on the RHS of $\tilde U_{rijk}$.
We now want to solve equations \eqref{eq:tildeUijkl}-\eqref{eq:tildeUij} for $\Phi_{rr}$, $\Phi_{ri}$ and $\Phi_{ij}$.

We will first decompose $\Phi_{ij}$ into a transverse traceless part ($\Phi^{TT}_{ij}$), a longitudinal traceless part ($\psi_i$) and a trace term, i.e.
\begin{equation}
    \Phi_{ij}=\Phi^{\text{TT}}_{ij}+\partial_i\psi_j+\partial_j\psi_i-\frac{2}{3}\delta_{ij}\partial_k\psi_k+\frac{1}{3}\delta_{ij}\Phi_{kk}\,.
\end{equation}
Using \eqref{eq:trafoPhiij} we can always choose a gauge such that 
\begin{equation}
    \Phi_{ij}=\Phi^{\text{TT}}_{ij}\,.
\end{equation}
The residual gauge transformations are those $\zeta_r$ and $\zeta_i$ that obey
\begin{eqnarray}
    \zeta_r & = & \frac{r}{3}\partial_i\zeta_i\,,\\
    0 & = & \partial_i\left(\partial_i\zeta_j+\partial_j\zeta_i-\frac{2}{3}\delta_{ij}\partial_k\zeta_k\right)=\partial^2\zeta_j+\frac{1}{3}\partial_j\partial_i\zeta_i\,.\label{eq:conditionzetai}
\end{eqnarray}
These act on the components of $\Phi_{\mu\nu}$ as
\begin{eqnarray}
    \delta\Phi_{rr} & = & \frac{4}{3}\partial_i\zeta_i+\frac{2}{3}r\partial_r\partial_i\zeta_i\,,\\
    \delta\Phi_{ri} & = & \frac{r}{3}\partial_i\partial_j\zeta_j+\partial_r\zeta_i+2r^{-1}\zeta_i\,,\label{eq:resgaugePhiir}\\
    \delta\Phi^{\text{TT}}_{ij} & = & \partial_i\zeta_j+\partial_j\zeta_i-\frac{2}{3}\delta_{ij}\partial_k\zeta_k\,,
\end{eqnarray}
where $\zeta_i$ obeys \eqref{eq:conditionzetai}.

In this gauge, equations \eqref{eq:tildeUijkl} to \eqref{eq:tildeUij} are
\begin{eqnarray}
    0 & = & \partial_i\partial_k\Phi^{\text{TT}}_{jl}+r^{-1}\delta_{ik}\left(\partial_j\Phi_{rl}+\partial_l\Phi_{rj}-\partial_r\Phi_{jl}^{\text{TT}}-2r^{-2}\Phi_{jl}^{\text{TT}}-r^{-2}\delta_{jl}\Phi_{rr}+E\delta_{jl}\right)\nonumber\\
    &&-(i\leftrightarrow j)-(k\leftrightarrow l)\,,\label{eq:ijkl}\\
    0 & = & \left(\partial_r+2r^{-1}\right)\left(\partial_j\Phi_{ik}^{\text{TT}}-\partial_k\Phi_{ij}^{\text{TT}}\right)-\partial_i\left(\partial_j\Phi_{rk}-\partial_k\Phi_{rj}\right)\nonumber\\
    &&+r^{-1}\left(\delta_{ik}\partial_j\Phi_{rr}-\delta_{ij}\partial_k\Phi_{rr}\right)\,,\label{eq:ijk}\\
    0 & = & \partial_r^2\Phi_{ij}^{\text{TT}}+3r^{-1}\partial_r\Phi_{ij}^{\text{TT}}-\left(\partial_r+r^{-1}\right)\left(\partial_i\Phi_{rj}+\partial_j\Phi_{ri}\right)+\partial_{i}\partial_j\Phi_{rr}\nonumber\\
    &&+\delta_{ij}\left(r^{-1}\partial_r\Phi_{rr}+2G\right)\,.\label{eq:ij}
\end{eqnarray}
Contracting \eqref{eq:ijkl} with $\delta_{ik}\delta_{jl}$ leads to
\begin{equation}
    E=r^{-2}\Phi_{rr}-\frac{2}{3}r^{-1}\partial_i\Phi_{ri}\,.
\end{equation}
Substituting this back into \eqref{eq:ijkl} and contracting with $\delta_{ik}$ gives us
\begin{equation}\label{eq:traceijkl}
    \partial^2\Phi_{jl}^{\text{TT}}-r^{-1}\partial_r\Phi_{jl}^{\text{TT}}-2r^{-2}\Phi_{jl}^{\text{TT}}+r^{-1}\left(\partial_j\Phi_{rl}+\partial_l\Phi_{rj}-\frac{2}{3}\delta_{jl}\partial_i\Phi_{ri}\right)=0\,.
\end{equation}
Using this \eqref{eq:ijkl} reduces to
\begin{equation}\label{eq:TTPhi}
    \partial_i\partial_k\Phi^{\text{TT}}_{jl}-\delta_{ik}\partial^2\Phi_{jl}^{\text{TT}}-(i\leftrightarrow j)-(k\leftrightarrow l)=0\,.
\end{equation}
Taking the divergence of \eqref{eq:traceijkl} gives
\begin{equation}\label{eq:EOMPhiir}
    \partial_j\left(\partial_j\Phi_{rl}+\partial_l\Phi_{rj}-\frac{2}{3}\delta_{jl}\partial_i\Phi_{ri}\right)=\partial^2\Phi_{rl}+\frac{1}{3}\partial_l\partial_j\Phi_{rj}=0\,.
\end{equation}
Contracting \eqref{eq:ijk} with $\delta_{ik}$ allows us to write
\begin{equation}
    \partial^2\Phi_{rj}-\partial_j\partial_i\Phi_{ri}+2r^{-1}\partial_j\Phi_{rr}=0\,.
\end{equation}
Taking the divergence of this equation we find
\begin{equation}\label{eq:harmPhirr}
    \partial^2\Phi_{rr}=0\,.
\end{equation}
Continuing this way we next contract \eqref{eq:ij} with $\delta_{ij}$ giving
\begin{equation}
    G(r)=\frac{1}{3}\left(\partial_r+r^{-1}\right)\partial_i\Phi_{ri}-\frac{1}{2}r^{-1}\partial_r\Phi_{rr}\,,
\end{equation}
where we used \eqref{eq:harmPhirr}.
Substituting this back into \eqref{eq:ij} we obtain
\begin{equation}
    \partial_r^2\Phi_{ij}^{\text{TT}}+3r^{-1}\partial_r\Phi_{ij}^{\text{TT}}-\left(\partial_r+r^{-1}\right)\left(\partial_i\Phi_{rj}+\partial_j\Phi_{ri}-\frac{2}{3}\delta_{ij}\partial_k\Phi_{rk}\right)+\partial_i\partial_j\Phi_{rr}=0\,.
\end{equation}

The residual gauge transformation \eqref{eq:resgaugePhiir} can be written as
\begin{equation}
    \delta\Phi_{ri}=\tilde\zeta_i\,,
\end{equation}
where 
\begin{equation}
    \tilde\zeta_i=\frac{r}{3}\partial_i\partial_j\zeta_j+\partial_r\zeta_i+2r^{-1}\zeta_i\,.
\end{equation}
The parameter $\tilde\zeta_i$ satisfies the same equation as $\zeta_i$, i.e.
\begin{equation}
    \partial^2\tilde\zeta_j+\frac{1}{3}\partial_j\partial_i\tilde\zeta_i=0\,.
\end{equation}
Other than this condition, $\tilde\zeta_i$ is arbitrary. Since $\Phi_{ri}$ obeys the same equation as $\tilde\zeta_i$, i.e. \eqref{eq:EOMPhiir} we have enough residual gauge freedom to set $\Phi_{ri}=0$. The residual gauge symmetry of setting $\Phi_{ri}=0$ is given by
\begin{equation}
    \tilde\zeta_i=\frac{r}{3}\partial_i\partial_j\zeta_j+\partial_r\zeta_i+2r^{-1}\zeta_i=0\,.
\end{equation}
This can be solved by
\begin{equation}
    \zeta_i=r^{-2}\varphi_i(x)-\frac{1}{6}\partial_i\partial_j \varphi_j\,,
\end{equation}
where $\varphi_i$ is independent of $r$ and obeys
\begin{equation}\label{eq:condvarphii}
    \partial^2 \varphi_i+\frac{1}{3}\partial_i\partial_j \varphi_j=0\,.
\end{equation}
Using $\Phi_{ri}=0$ it follows that the remaining components of $\Phi_{\mu\nu}$ satisfy
\begin{eqnarray}
 \left(\partial_r+2r^{-1}\right)\left(\partial_j\Phi_{ik}^{\text{TT}}-\partial_k\Phi_{ij}^{\text{TT}}\right) & = & 0\,,\label{eq:cond1}\\
    \partial^2\Phi_{jl}^{\text{TT}}-r^{-1}\partial_r\Phi_{jl}^{\text{TT}}-2r^{-2}\Phi_{jl}^{\text{TT}} & = & 0\,,\\
    \partial_r^2\Phi_{ij}^{\text{TT}}+3r^{-1}\partial_r\Phi_{ij}^{\text{TT}}& = & 0\,,\label{eq:cond3}
\end{eqnarray}
as well as \eqref{eq:TTPhi} and
\begin{equation}
   \partial_j\Phi_{rr}=0\,.
\end{equation}
Equations \eqref{eq:cond1}--\eqref{eq:cond3} are solved by
\begin{equation}
    \Phi_{ij}^{\text{TT}}=r^{-2}\tilde\Phi_{ij}^{\text{TT}}(x)+\frac{1}{2}\partial^2\tilde\Phi_{ij}^{\text{TT}}\,,
\end{equation}
where $\tilde\Phi_{ij}^{\text{TT}}(x)$ is independent of $r$ and obeys
\begin{equation}\label{eq:curlpartial^2Phi}
    \partial_j \partial^2\tilde\Phi_{ik}^{\text{TT}}-\partial_k \partial^2\tilde\Phi_{ij}^{\text{TT}} = 0\,.
\end{equation}
The residual gauge symmetry acting on $\tilde\Phi_{ij}^{\text{TT}}(x)$ is
\begin{equation}
    \delta\tilde\Phi_{ij}^{\text{TT}}(x)=\partial_i\varphi_j+\partial_j\varphi_i-\frac{2}{3}\delta_{ij}\partial_k\varphi_k\,,
\end{equation}
where $\varphi_i$ obeys \eqref{eq:condvarphii}.
Using this residual gauge transformation as well as \eqref{eq:curlpartial^2Phi} we can always fix the gauge further by setting\footnote{This follows from the fact that \eqref{eq:curlpartial^2Phi} tells us that $\partial^2\tilde\Phi_{ik}^{\text{TT}}=\partial_i\partial_k H$ where $H$ is harmonic in order that the RHS is transverse traceless. Furthermore, $\partial_i\varphi_i$ is also harmonic but further arbitrary. Hence, since we have
\begin{equation}
    \delta\partial^2\tilde\Phi_{ij}^{\text{TT}}=-\frac{2}{3}\partial_i\partial_j\left(\partial_k\varphi_k\right)\,,
\end{equation}
it follows that we can make the further gauge choice \eqref{eq:gaugechoicepartial^2Phi}.}
\begin{equation}\label{eq:gaugechoicepartial^2Phi}
    \partial^2\tilde\Phi_{ik}^{\text{TT}}=0\,.
\end{equation}
We are now finally left with 
\begin{equation}
    \Phi_{ij}^{\text{TT}}=r^{-2}\tilde\Phi_{ij}^{\text{TT}}(x)\,,
\end{equation}
where $\tilde\Phi_{ij}^{\text{TT}}(x)$ obeys 
\begin{equation}\label{eq:finaleqPhiTT}
    \partial_i\partial_k\tilde\Phi^{\text{TT}}_{jl}-\partial_j\partial_k\tilde\Phi^{\text{TT}}_{il}
    -\partial_i\partial_l\tilde\Phi^{\text{TT}}_{jk}+\partial_j\partial_l\tilde\Phi^{\text{TT}}_{ik}
    =0\,,
\end{equation}
as follows from \eqref{eq:TTPhi}. The trace of this equation implies \eqref{eq:gaugechoicepartial^2Phi}. The residual gauge transformations acting on $\tilde\Phi_{ij}^{\text{TT}}(x)$ are of the form\footnote{One could relax this slightly by only demanding that $\partial_i\partial_j\left(\partial_k\varphi_k\right)=0$ together with \eqref{eq:condvarphii}, but this will not be necessary. The transformations that we are leaving out here can anyway be recovered by computing the residual gauge invariance of the final result given in \eqref{eq:finalresult}.}
\begin{equation}\label{eq:resgaugePhiij}
    \delta \tilde\Phi_{ij}^{\text{TT}}(x)=\partial_i\varphi_j^\text{T}+\partial_j\varphi_i^\text{T}\,,
\end{equation}
where $\varphi_j^\text{T}$ is transverse and harmonic. We will finally show that we have enough residual gauge freedom to use \eqref{eq:finaleqPhiTT} to set $\tilde\Phi_{ij}^{\text{TT}}=0$. To show this we write \eqref{eq:finaleqPhiTT} as
\begin{equation}\label{eq:dQ}
    \partial_k Q_{ijl}-\partial_l Q_{ijk}=0\,,
\end{equation}
where 
\begin{equation}\label{eq:Q}
    Q_{ijl}=\partial_i\tilde\Phi_{jl}^{\text{TT}}-\partial_j\tilde\Phi_{il}^{\text{TT}}\,.
\end{equation}
The object $Q_{ijl}$ is traceless and has a hook symmetric Young tableaux, i.e. it is antisymmetric in its first two indices and the totally antisymmetric part, $Q_{[ijl]}$, is zero. Equation \eqref{eq:dQ} is an integrability condition that is locally solved by $Q_{ijl}=\partial_l Q_{ij}$ where $Q_{ij}=-Q_{ji}$ and where $Q_{[ijl]}=\partial_{[l} Q_{ij]}=0$ tells us that $Q_{ij}=\partial_i Q_j-\partial_j Q_i$. Using \eqref{eq:Q} we see that 
\begin{equation}
    \tilde\Phi_{jl}^{\text{TT}}=\partial_j Q_l+\partial_l Q_j\,.
\end{equation}
Transversality and tracelessness of the LHS then tells us that $Q_i$ must be transverse and harmonic, and can thus be gauged away using \eqref{eq:resgaugePhiij}.

\subsubsection{Constructing planar symmetric $m_\mu$}
Next, we turn our attention to the gauge field $m_\mu$. The total gauge transformation acting on $m_\mu$ including diffeomorphisms is
\begin{equation}
    \delta m_\mu=\mathcal{L}_K m_\mu+\mathcal{L}_\zeta\tau_\mu\,.
\end{equation}
The vector $K^\mu$ is the generator of a symmetry if this total variation is zero. For the spatial components of $m_\mu$ we have a field strength 
\eqref{eq:defF} which is invariant under NLO diffeomorphisms for TTNC geometries \eqref{eq:trafohhF}.

For our LO geometry the spatial components are $F_{ij}$ and $F_{ri}$ given by
\begin{eqnarray}
    F_{ij} & = & \partial_i m_j-\partial_j m_i\,,\\
    F_{ri} & = & \partial_r m_i-\partial_i m_r+r^{-1}m_i=0\,.
\end{eqnarray}
Similar to our treatment of the  $\Phi_{ab}$ field, we want to impose the symmetries in \eqref{eq:Killingvec}. This means that we have
\begin{equation}\label{eq:F}
    \mathcal{L}_K F_{ij}=0\,,\qquad \mathcal{L}_K F_{ri}=0\,.
\end{equation}
It follows that $F_{ij}=0$ and $F_{ir}=0$ since there are no $SO(3)$ invariants that could appear on the RHS and respect the antisymmetric index structure.

The gauge transformation of $m_\mu$ can be written as
\begin{equation}
    \delta m_\mu=\partial_\mu\Lambda-a_\mu\Lambda+\tau_\mu a_\rho\zeta^\rho\,,
\end{equation}
which in our case leads to
\begin{eqnarray}
    \delta m_i & = & \partial_i\Lambda\,,\\
    \delta m_r & = & \partial_r\Lambda+r^{-1}\Lambda\,.
\end{eqnarray}
Using $F_{ij}=F_{ri}=0$ we can always pick a gauge in which
\begin{equation}
    m_i=0=m_r\,.
\end{equation}
The residual gauge transformations are given by
\begin{equation}\label{eq:resLambda}
    \Lambda = r^{-1}f(t)\,,
\end{equation}
where $f(t)$ is any function time.

This leaves us with the time component $m_t$. In order for this to respect the Killing symmetries we require that the Lie derivative of $m_t$ is equal to (minus) a gauge symmetry so that the total gauge transformation of $m_t$ is zero. Explicitly, this leads to
\begin{equation}
    \mathcal{L}_K m_t=-\partial_t\Lambda-\tau_t a_\rho\zeta^\rho=-\partial_t\Lambda+l^{-1}\zeta_r\,,
\end{equation}
where the RHS consists of residual gauge transformations of the previous gauge choices. The residual gauge transformations are not very powerful due to all the previous choices that we made. Hence, we can make a further gauge choice by imposing the condition $\mathcal{L}_K m_t=0$ which leads to 
\begin{equation}
    m_t=m_t(r)\,.
\end{equation}

We thus conclude that imposing boundary time and space translation as well as boundary rotation invariance leads to
\begin{eqnarray}
    &&\Phi_{ij}=0\,,\qquad\Phi_{ri}=0\,,\qquad\Phi_{rr}=\Phi_{rr}(r)\,,\label{eq:finalresult}\\
    && m_i=0\,,\qquad m_r=0\,,\qquad m_t=m_t(r)\,.
\end{eqnarray}

\section{Non-relativistic expansions of asymptotically AdS spacetimes} \label{App:NRExpansionOfAsympAdS}

\subsection{AdS black brane}\label{app:AdSBB}
The metric of the AdS$_{d+2}$ black brane is given by 
\begin{align}\label{eq:BB}
    ds^2=& - \frac{R^2}{l^2} f(R) c^2 dt^2 + \frac{l^2}{R^2}f^{-1}(R) dR^2  + \frac{R^2}{l^2} d\vec{x}^2, \\
    f(R) =& 1 - \frac{1}{(bR)^{d+1}}
\end{align}
where $d\vec{x} = (dx^1,dx^2,...,dx^d)$ so that $D=2+d$. We want to do a $\frac{1}{c^2}$-expansion of this metric but in order to do that we must first reinstate factors of $c$ in $f(R)$.  

We know that the temperature and entropy density of the AdS black brane is related to its energy density and pressure through the thermodynamic relation
\begin{align}
    \mathcal{E} + P = sT, \label{eq:thermorel}
\end{align}
where $\mathcal{E}$ is the energy density, $P$ the pressure, $s$ the entropy density and $T$ the temperature. 
\begin{align}
    s &= \frac{k_B c^3}{4 G \hbar (bl)^{d}}. \label{eq:Ent}
    \\
    T_{BH} &= \frac{d+1}{4\pi b l^2} \frac{c\hbar}{k_B}, \label{eq:Temp}
\end{align}
The equation of state is  $\mathcal{E}=dP$. Using this along with \eqref{eq:thermorel}-\eqref{eq:Temp} we get
\begin{equation}\label{eq:Etob}
    \mathcal{E}=d\frac{c^4}{16\pi G}l^{-(d+2)}b^{-(d+1)}=mc^2\,,
\end{equation}
where we defined the mass density by $\mathcal{E}=mc^2$. Using this to eliminate $b$ from the function $f$ we get
\begin{equation}
    f=1-\frac{16\pi G}{(D-2)c^2}l^D m R^{-(D-1)}\,.
\end{equation}

To introduce the fluid velocity $v^i$ we perform a boundary Lorentz boost transformation of the AdS black brane solution leading to
\begin{align}\label{eq:boostedBB}
      ds^2= \frac{l^2}{R^2}f^{-1}(R) dR^2 + \frac{R^2}{l^2}\big[ - f(R) \frac{1}{c^2}u_A u_B   + P_{AB} \big] dx^A dx^B\,,
\end{align}
where $A,B$ runs over $t$ and the spatial directions $x^i$, and where
\begin{align}
    u_A &= \frac{1}{\sqrt{1-v^2/c^2}} (-c^2, v^i)\,,
    \\
    P_{AB} &= \frac{1}{c^2}u_A u_B + \eta_{AB}\,.
\end{align}
The $1/c^2$ expansion (assuming that $l$ does not depend on $c$) then tells us that 
\begin{eqnarray}
    \tau_\mu dx^\mu & = & \frac{R}{l}dt\,,\\
    h_{\mu\nu}dx^\mu dx^\nu & = & \frac{l^2}{R^2}dR^2+\frac{R^2}{l^2}d\vec x^2\,,\\
    m_\mu dx^\mu & = & -\frac{8\pi G}{D-2}ml^{D-1}R^{-(D-2)}\,.\label{eq:mt}
\end{eqnarray}

\subsection{Non-relativistic boundary energy-momentum tensor}\label{app:bdryEMT}

In this subsection we fix $D=5$. Doing a Fefferman-Graham expansion around $AdS_5$ we get 
\begin{align}
    ds^2 = l^2 \frac{dr^2}{r^2} + \frac{l^2}{r^2} \big( g^{(0)}_{AB} + r^2 g^{(2)}_{AB} + r^4 g^{(4)}_{AB} +...\big) dx^A dx^B, \label{eq:BoundaryMetric}
\end{align}
where again $A,B=(t,i)$. If we assume that $g^{(0)}_{AB} = \eta_{AB}$ it follows that $g^{(2)}_{AB} = 0$ and
then we know that the boundary stress tensor is given by
\begin{align}\label{eq:bdryEMT}
    T_{\ B}^A = \frac{l^3c^4}{4\pi G} \eta^{AC}g^{(4)}_{CB},
\end{align}
where the factors of $G$, $l$ and $c$ were found by dimensional analysis by demanding that $T^t{}_t$ has dimensions of energy density in $4$ spacetime boundary dimensions. Note that $G$ is Newton's constant in $5$ bulk dimensions. For the AdS black brane with $D=5$ we get from \eqref{eq:bdryEMT} that $T^t{}_t=-\mathcal{E}$. To see this we take \eqref{eq:BB} and change coordinates according to 
\begin{equation}\label{eq:toFG}
    R^2=\frac{l^4}{r^2}\left(1+\frac{r^4}{4l^8b^4}\right)\,.
\end{equation}
This puts \eqref{eq:BB} in FG form. Using \eqref{eq:bdryEMT} and \eqref{eq:Etob} then gives $T^t{}_t=-\mathcal{E}$.

Due to Lorentz symmetry we have
\begin{equation}
    \eta_{C[A}T^C{}_{B]}=0\,.
\end{equation}
From this we conclude that
\begin{equation}\label{eq:LBWI}
    c^2 T^t{}_i=-T^i{}_t\,.
\end{equation}
When the boundary metric is the Minkowski metric we can write the FG expansion as
\begin{eqnarray}
    ds^2 & = & \frac{l^2}{r^2}dr^2+\frac{l^2}{r^2}\left(-c^2dt^2+d\vec x^2\right)+\frac{4\pi G}{l}r^2\left[-c^{-2}T^t{}_t dt^2\right.\nonumber\\
    &&\left.+2c^{-4}T^i{}_t dtdx^i+c^{-4}T^i{}_j dx^i dx^j\right]+\mathcal{O}{(r^4)}\,, \label{eq:EMTinMet}
\end{eqnarray}
where we used \eqref{eq:LBWI}.

Due to $T^t{}_t=-\mathcal{E}=-mc^2$, where we treat $m$ as independent of $c$, and because an energy-momentum tensor is a $(1,1)$ tensor we expand 
\begin{align}
T^A_{\ B} = c^2\mathcal{T}^A_{\ B} +\overset{(0)}{\mathcal{T}}{}^A_{\ B}+\mathcal{O}(c^{-2})\,.
\end{align}
This along with \eqref{eq:EMTinMet} leads to the following identification between the EMT and the $1/c^2$-expansion of the metric
\begin{eqnarray}
    \tau_\mu dx^\mu & = & \frac{l}{r}dt\,,\\
    \bar h_{\mu\nu}dx^\mu dx^\nu & = & \frac{l^2}{r^2}\left(dr^2+d\vec x^2\right)-\frac{4\pi G}{l}r^2\mathcal{T}^t{}_t dt^2-\frac{4\pi G}{l}r^2\mathcal{T}^t{}_i dtdx^i+\mathcal{O}(r^4)\,, \label{eq:EMTinNRMet1}\\
    \bar \Phi_{\mu\nu}dx^\mu dx^\nu & = & \frac{4\pi G}{l}r^2\left(-\overset{(0)}{\mathcal{T}}{}^t{}_{t}dt^2-\overset{(0)}{\mathcal{T}}{}^t{}_{i}dt dx^i+\mathcal{T}^i{}_t dt dx^i+\mathcal{T}^i{}_j dx^i dx^j\right)+\mathcal{O}(r^4)\,, \label{eq:EMTinNRMet2}
\end{eqnarray}
where the definition of $\bar{h}_{\mu \nu}$ and $\bar{\Phi}_{\mu \nu}$ is given in \eqref{eq:hbar} and \eqref{eq:Phibar}.

If we use \eqref{eq:LBWI} then we get
\begin{equation}
    \mathcal{T}^t{}_i=0\,,\qquad -\overset{(0)}{\mathcal{T}}{}^t{}_{i}=\mathcal{T}^i{}_t\,,
\end{equation}
so that 
\begin{align}
    &\mathcal{T}^t_{\ t} = - \frac{l}{4\pi G} \bar{h}_{tt}^{[2]},
    \qquad
    \mathcal{T}^i_{\ j} = \frac{l}{4\pi G } \bar{\Phi}_{ij}^{[2]}
    \qquad
    \mathcal{T}^i_{\ t} = \frac{l}{4\pi G } \bar{\Phi}_{it}^{[2]}\,,
\end{align}
where the $[2]$ superscript indicates the coefficient of $r^2$.

Finally, we transform \eqref{eq:boostedBB} using the coordinate transformation given by \eqref{eq:toFG}. We then $1/c^2$ expand and obtain the result 
\begin{subequations}
    \begin{align}
    \tau_\mu dx^\mu =& \frac{l}{r} dt\,,\\
        \bar{h}_{\mu \nu} dx^\mu dx^\nu =& \frac{l^2}{r^2} \big( dr^2 + d \vec{x}^2 \big) + \frac{4\pi G m }{l} r^2  dt^2 + \mathcal{O}(r^5)\,,
        \\
        \bar{\Phi}_{\mu \nu} dx^\mu dx^\nu =& \frac{4 \pi Gm  r^2}{l} \left( \frac{1}{3} d\vec{x}^2 +\frac{4}{3} v^2 dt^2 - \frac{8}{3} v_i dx^i dt\right)  +\mathcal{O}(r^4)\,.
    \end{align}
\end{subequations}
Using this we find 
\begin{subequations}
    \begin{align}
        \mathcal{T}^t_{\ t} =  -m\,,
        \qquad
         \mathcal{T}^i_{\ t}= - \frac{4 m}{3} v^i\,,
        \qquad
        \mathcal{T}^t_{\ i} =0\,,\qquad
        \mathcal{T}^i_{\ j} = \frac{m}{3}\,.
    \end{align}
\end{subequations}

\bibliographystyle{utphys} 
\bibliography{bibl} 

\end{document}